\documentclass[electronics,article,submit,pdftex,moreauthors]{Definitions/mdpi}

\firstpage{1}
\pubvolume{1}
\issuenum{1}
\articlenumber{0}
\pubyear{2026}
\copyrightyear{2026}
\datereceived{ }
\daterevised{ }
\dateaccepted{ }
\datepublished{ }

\newcommand{\thead}[1]{\begin{tabular}[c]{@{}c@{}}#1\end{tabular}}

\newcommand{\rheadcell}[1]{\begin{tabular}[c]{@{}r@{}}#1\end{tabular}}

\Title{INC-SRv6: In-Network Route Exploration and Fast Failover Using Segment Routing Headers in Dynamic UAV Networks}

\Author{Zhiyuan Ren \texorpdfstring{$^{1,*}$}{} and Zhao Li \texorpdfstring{$^{2}$}{}}

\AuthorNames{Zhiyuan Ren and Zhao Li}

\address[1]{%
$^{1}$ \quad School of Telecommunications Engineering, Xidian University, Xi'an 710071, China; zyren@xidian.edu.cn\\
$^{2}$ \quad Shaanxi Academy of Aerospace Technology Application Co., Ltd., No. 150 Weiqu West Street, Chang'an District, Xi'an 710100, Shaanxi, China; lizhao\_829@qq.com}

\corres{Correspondence: zyren@xidian.edu.cn}

\usepackage{algpseudocode}
\makeatletter
\renewcommand{\theHALG@line}{\thealgorithm.\arabic{ALG@line}}
\makeatother
\newenvironment{algorithmblock}[1]{%
  \par\medskip\noindent\begin{minipage}{\linewidth}
  \refstepcounter{algorithm}\small
  \textbf{Algorithm \thealgorithm. #1}\par\smallskip\hrule\smallskip
}{\smallskip\hrule\end{minipage}\par\medskip}

\abstract{Unmanned aerial vehicle (UAV) networks require a network provisioning mechanism that guarantees delivery and latency quality of service (QoS) under dynamic conditions in next-generation airborne mesh networks by continuously learning and repairing multi-hop paths under rapid link turnover without flooding the control plane. This paper presents INC-SRv6, an in-network computing routing mechanism that repurposes the IPv6 Segment Routing Header (SRH) and its \texttt{Segments Left} (\texttt{SL}) pointer into a self-learning, fast-failover distributed control plane. Rather than treating the SRH as a static source-committed vector, INC-SRv6 uses the active \texttt{SL} pointer to partition the segment list in flight: slots ahead of \texttt{SL} provide capacity for hop-by-hop next-hop recording, while the suffix behind \texttt{SL} serves as an in-band multi-destination reachability vector. Through this dual-zone header discipline, transit routers extract distance-weighted reachability updates from arriving suffixes, gate admission by single-pass confidence with bounded binary probing, advance or reverse \texttt{SL} deterministically, and restore pre-transmission SRH state for local failover upon link loss. Simulations across 200-node 3D UAV networks show that under high-speed flight and repeated \texorpdfstring{$20\%$}{20\%} relay outages, INC-SRv6 improves packet delivery over reactive routing (\texorpdfstring{$92.12\%$}{92.12\%} versus \texorpdfstring{$89.12\%$}{89.12\%}) while cutting 95th-percentile tail delay from \texorpdfstring{$121.26$}{121.26} to \texorpdfstring{$8.84$}{8.84}~ms and control packets by \texorpdfstring{$35.8\%$}{35.8\%}, with a 4-byte compact \texttt{uSID} stride model reducing SRv6 control bytes by over \texorpdfstring{$52\%$}{52\%}.}

\keyword{Segment Routing over IPv6 (SRv6); Segment Routing Header (SRH); Segments Left (\texttt{SL}) pointer; in-network computing (INC); unmanned aerial vehicle (UAV) networks; distributed route exploration; fast failover; compressed SRv6 SID (\texttt{uSID}); network provisioning; quality of service (QoS); next-generation networks}

\renewcommand{\linenumbers}{}
\begin{document}

\section{Introduction}
\label{sec:intro}

Unmanned aerial vehicle (UAV) swarms provide rapidly deployable, infrastructure-less communication backbones for emergency response, disaster relief, and wide-area sensing~\cite{gupta2016survey,jiang2018routing,lakew2020routing,yao2022uavfleet}. In a large-scale multi-hop aerial mesh, end-to-end connectivity depends on intermediate UAV relays whose geometric positions and operational states change continuously during flight. Even when a source--destination pair remains topologically connected through alternative trajectories, high-speed maneuvers or sudden relay outages can sever active forwarding links on the sub-second timescale. Sustaining reliable low-latency communication and statistical quality-of-service (QoS) guarantees in such dynamic non-terrestrial environments~\cite{wang2024statqos} requires a routing architecture that discovers feasible multi-hop paths without broadcast storms, refreshes local forwarding state continuously as contacts evolve, and repairs broken next hops locally without stalling active flows.

Existing routing paradigms struggle to satisfy these requirements simultaneously~\cite{oudi2023fanet,alam2022qlearning,zhang2023qlearninguav}. Proactive link-state protocols disseminate periodic topology advertisements across the swarm, incurring heavy background overhead while leaving computed paths vulnerable to topological staleness between update intervals. Reactive distance-vector protocols suppress idle overhead by discovering routes on demand, but rely on network-wide request flooding and upstream route-error notifications (\texttt{RERR}) that stall source transmission during link breaks and inflate tail latency into the hundred-millisecond regime. Classical source routing records explicit route vectors during discovery, yet commits each data packet at the source to a fixed end-to-end path that frequently breaks in flight under high mobility. Meanwhile, online reinforcement learning schemes using one-hop delay feedback propagate multi-hop reachability too slowly to track rapid 3D link turnover, whereas continuous multi-agent swarm probing generates prohibitive control packet volumes.

To overcome this fundamental tension, we look to a standardized protocol header already present in modern IPv6 data planes: the Segment Routing Header (SRH, RFC~8754) in Segment Routing over IPv6 (SRv6)~\cite{rfc8402,rfc8754,rfc8986,ventre2020srv6,polverini2025srv6detnet}. In conventional SRv6 deployments, the SRH is treated as a static, read-only instruction list computed by a centralized controller or link-state database~\cite{rovira2021sdnuav,cianfrani2018sdnsrv6}: the source preloads an explicit sequence of 128-bit Segment Identifiers (SIDs) into \texttt{SRH.segment\_list}, and transit routers merely decrement the \texttt{Segments Left} (\texttt{SL}) pointer to read the next hop. Under rapid UAV churn, such source-locked segment lists quickly become obsolete.

The central insight of this paper is that the standard RFC~8754 SRH structure, through the interaction between \texttt{SRH.segment\_list} and the \texttt{SL} pointer, can be repurposed into a dynamic, in-flight state register for in-network routing computation~\cite{sapio2017innetwork,ports2019when}. While preserving the standard forwarding invariant $\texttt{IPv6.DA} = \texttt{SRH.segment\_list}[\texttt{SL}]$ at every hop, the active \texttt{SL} cursor naturally partitions the segment list into two functional regions during packet traversal. Slots ahead of \texttt{SL} ($\texttt{segment\_list}[1..\texttt{SL}-1]$) provide pre-allocated capacity where transit routers dynamically record selected next-hop SIDs as they decrement \texttt{SL}, rather than following a rigid source-committed route. Simultaneously, the traversed suffix behind \texttt{SL} ($\texttt{segment\_list}[\texttt{SL}+1..\texttt{LastEntry}]$) preserves the exact sequence of visited routers. Every arriving packet therefore carries an in-band multi-destination reachability vector: for any suffix index $k > \texttt{SL}$, the receiving router observes that node $\texttt{segment\_list}[k]$ is reachable via neighbor $\texttt{segment\_list}[\texttt{SL}+1]$ at an exact hop distance of $h = k - \texttt{SL}$.

Building on this dual-zone SRH representation, we propose \textbf{INC-SRv6}, an in-network computing mechanism that unifies distributed route exploration, multi-destination learning, loop-free forwarding, and local fast failover within the SRv6 data plane. In INC-SRv6, transit routers continuously extract distance-attenuated reachability updates for all intermediate nodes in arriving packet suffixes, amortizing discovery traffic across intersecting flows and pre-populating backup next hops before links break. When a source lacks a confident route, it launches a bounded binary exploration wave whose branches exclude nodes already present in their SRH suffix to guarantee loop freedom, while the destination reverses the compacted SRH array once so that the acknowledgment retraces the discovered path strictly by decrementing \texttt{SL} without routing lookups. During subsequent data forwarding, routers make fresh hop-by-hop decisions using locally learned weights; if an outbound link fails just before transmission, the detecting router rolls back the packet's pre-hop SRH state to expunge the untraversed hop and immediately rewrites an alternate loop-free next hop, avoiding upstream route-error propagation and source rediscovery stalls.

The main contributions of this work are summarized as follows:
\begin{enumerate}[leftmargin=1.8em]
    \item \textbf{Dual-zone SRH state representation for in-network routing:} We formulate a unified SRH state representation and pointer discipline that transforms \texttt{SRH.segment\_list} and \texttt{SL} from a static source-route header into a dual-zone in-band telemetry and control structure supporting hop-by-hop next-hop recording, multi-destination suffix extraction, loop-free candidate filtering, and lookup-free reverse acknowledgments.
    \item \textbf{Closed-loop exploration, confidence gating, and local failover:} We design a self-learning forwarding mechanism that couples single-pass confidence-calibrated reinforcement ($\alpha(h) > \theta$), least-probed binary exploration cloning ($\le 4$ leaves per generation), and pre-transmission SRH snapshot rollback ($\sigma$) for immediate loop-free local repair.
    \item \textbf{Large-scale 3D UAV network evaluation:} Across $1,100$ simulation runs spanning $10$ mobility and fault scenarios in $200$-node 3D UAV networks, we benchmark INC-SRv6 against five protocol families (\texttt{OSPF-SRv6}, \texttt{OLSR-L3}, \texttt{AODV-L3}, \texttt{Q-Routing-L3}, and \texttt{AntHocNet-L3}), four mechanism ablations, and a 4-byte compact \texttt{uSID} variant, quantifying the impact of each design component and demonstrating a $52.1\%$--$54.7\%$ control-byte reduction via 4-byte compact \texttt{uSID} stride encoding.
\end{enumerate}

The remainder of this paper is organized as follows. Section~\ref{sec:related} reviews related work. Section~\ref{sec:model} defines the 3D UAV network model and programmable SRH state representation. Section~\ref{sec:workflow} details the INC-SRv6 processing workflow. Section~\ref{sec:evaluation} presents the simulation evaluation, ablation study, and CPU microbenchmark. Section~\ref{sec:conclusion} concludes the paper.

\section{Related Work}
\label{sec:related}

Table~\ref{tab:related} summarizes the routing state representations, discovery mechanisms, and failure recovery strategies of representative routing protocols compared with INC-SRv6.

\begin{table}[H]
\caption{Comparison of routing-state representations, discovery mechanisms, and failure handling strategies across protocols. Baseline engineering parameters are specified in Section~\ref{sec:evaluation}.}
\label{tab:related}
\centering\footnotesize
\setlength{\tabcolsep}{3pt}
\begin{tabularx}{\linewidth}{@{}>{\raggedright\arraybackslash}p{1.65cm}LLL@{}}
\toprule
\textbf{Protocol} & \textbf{Routing state} & \textbf{Discovery / updates} & \textbf{Failure handling} \\
\midrule
OSPF-SRv6 & Local link-state database; source SRH & Periodic LSA dissemination & Local alternate from cached link-state paths \\
OLSR-L3 & Two-hop and MPR topology & HELLO and MPR-relayed TC & Neighbor expiry and route recomputation \\
AODV-L3 & Scalar next-hop and precursor entries & RREQ flood; reverse-hop RREP & RERR and source rediscovery \\
DSR & Cached route vectors; source-route header & RREQ flood; route-carrying RREP & Cached-route packet salvaging \\
Q-Routing-L3 & Per-destination/neighbor delay estimates & One-hop feedback & Next-best eligible neighbor \\
AntHocNet-L3 & Pheromone tables and recorded agent paths & Reactive/proactive probes and diffusion & Local fallback and renewed exploration \\
INC-SRv6 & Multi-destination weights from SRH suffixes & Bounded cloning; reverse-path ACK & Pre-hop restoration and local next-hop selection \\
\bottomrule
\end{tabularx}
\end{table}

\subsection{SRv6 Network Programming and Fast Reroute}
Segment Routing over IPv6 (SRv6) encodes forwarding instructions as an ordered list of 128-bit Segment Identifiers (SIDs) inside the Segment Routing Header (SRH, Routing Type~4), using the \texttt{Segments Left} (\texttt{SL}) pointer to index the active segment~\cite{rfc8402,rfc8754,rfc8986,abdelsalam2018srev6}. Beyond static source-routed traffic engineering and service function chain embedding~\cite{cianfrani2018sdnsrv6,ventre2020srv6,zheng2023sfc}, SRv6 Network Programming allows transit routers to execute lightweight header inspection, in-band performance monitoring, and state manipulation in eBPF or P4 pipelines~\cite{lebek2021ebpf,scarpitta2024srv6mon}, while recent IETF standards and micro-SID architectures specify compact SID addressing~\cite{rfc9602,tulumello2023usid} and compressed segment-list encodings (\texttt{uSID} / \texttt{NEXT-C-SID})~\cite{rfc9800} to reduce 128-bit header overhead.

For fault recovery and deterministic low-latency delivery~\cite{polverini2025srv6detnet}, IP Fast Reroute mechanisms such as Loop-Free Alternates (LFA)~\cite{rfc5286}, Topology Independent LFA (TI-LFA)~\cite{rfc9855}, and SDN-assisted UAV architectures~\cite{rovira2021sdnuav} precompute backup repair paths from a Link-State Database (LSDB) or centralized controller. However, under rapid 3D UAV mobility and simultaneous multi-relay churn, periodic Link-State Advertisements (LSAs) lag behind sub-second link transitions, causing source-computed SRHs and precomputed TI-LFA repair paths to steer packets into broken links. INC-SRv6 eliminates this dependency on global link-state synchronization by repurposing the SRH itself as an in-flight path-recording and learning structure, updating next-hop weights directly from arriving suffixes and repairing broken links locally.

\subsection{Routing in Dynamic UAV Networks}
Flying Ad Hoc Network (FANET) routing protocols span proactive, reactive, and learning-based designs~\cite{gupta2016survey,jiang2018routing,lakew2020routing,oudi2023fanet}. Proactive protocols such as OLSR~\cite{rfc3626} use Multipoint Relays (MPRs) to disseminate Topology Control (\texttt{TC}) messages; while providing immediate lookups in quasi-static topologies, periodic \texttt{TC} snapshots suffer severe blackholing when links change faster than the broadcast interval. Reactive protocols such as AODV~\cite{rfc3561} discover scalar routes on demand via flooded Route Requests (\texttt{RREQ}) and unicast Route Replies (\texttt{RREP}), using upstream Route Errors (\texttt{RERR}) upon link failure to trigger fresh source flooding. Under repeated relay churn, this drop-invalidate-reflood cycle introduces recurrent discovery stalls that inflate tail latency by an order of magnitude.

Dynamic Source Routing (DSR)~\cite{rfc4728} records traversed node addresses during discovery for route caching and salvaging, but commits each data packet at the source to a fixed end-to-end route vector that frequently breaks in flight as UAV speed and path length grow. INC-SRv6 differs fundamentally from DSR: (i)~rather than locking data packets to source-selected vectors, it converts observed SRH suffixes into normalized per-destination weight tables for fresh hop-by-hop decisions; (ii)~it bounds unconfident exploration to a binary tree of at most four leaves rather than network-wide flooding; and (iii)~Stage~4 restores the pre-transmission SRH snapshot before local rerouting so failed hops never pollute the recorded suffix.

Reinforcement learning and swarm-intelligence protocols replace rigid tables with adaptive estimates~\cite{boyan1993packet,dicaro2005anthocnet,da2021qfanet,alam2022qlearning,zhang2023qlearninguav,qiu2024marlfanet}. Q-Routing and recent multi-agent FANET extensions~\cite{boyan1993packet,da2021qfanet,alam2022qlearning,zhang2023qlearninguav,qiu2024marlfanet} update neighbor quality estimates via one-hop feedback, which bootstraps multi-hop reachability by only a single hop per transmission and converges too slowly under 3D mobility without external geographic beacons. AntHocNet~\cite{dicaro2005anthocnet} accelerates convergence via reactive/proactive ants and neighbor pheromone diffusion, but incurs millions of control packets in 200-node swarms. INC-SRv6 achieves rapid multi-hop convergence without swarm flooding by extracting up to $H_{\max}$ destination updates per arriving SRH suffix and calibrating its single-pass reinforcement step so that one validated acknowledgment immediately opens the source confidence gate.

\subsection{In-Network Computing and Path-Carried State}
In-network computing (INC) shifts state processing and inference into forwarding elements~\cite{sapio2017innetwork,ports2019when,zhang2024in3}, while In-band Network Telemetry (INT) and its low-overhead variants~\cite{kim2015int,qian2024offsetint} append device state to packets for passive monitoring. INC-SRv6 turns this in-band state paradigm into an active, closed-loop routing control plane built entirely within standard RFC~8754 SRH fields: every packet's traversed SRH suffix acts as a multi-destination reachability vector, and each router executes a four-stage pipeline that learns from the suffix, gates or branches outbound traffic, updates the \texttt{SL} pointer, and repairs local link failures in place.

\section{System Model and SRv6 Header State Representation}
\label{sec:model}

\subsection{3D Dynamic UAV Network Pure-L3 Model}
\begin{figure}[H]
    \centering
    \includegraphics[width=0.98\textwidth]{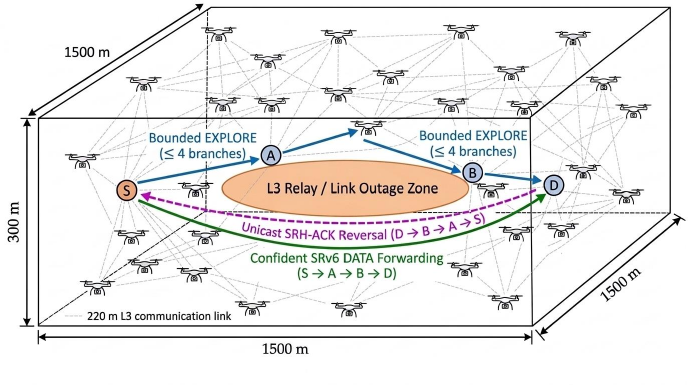}
    \caption{\textbf{3D dynamic UAV mesh network scenario ($N=200$ UAVs in $1500\times1500\times300$~m airspace, $R=220$~m L3 contact range) experiencing a central L3 relay/link outage zone.} Bounded \texttt{EXPLORE} packets ($\texttt{split\_rem}=2$, solid blue arrows, $\le 4$ leaf branches) detour around the outage while recording traversed SIDs into \texttt{SRH.segment\_list[SL]}. Destination $D$ returns a deterministic unicast \texttt{SRH-ACK} (dashed magenta arrows) along \texttt{IPv6.DA = SRH.segment\_list[SL]}, and high-confidence \texttt{DATA} packets (solid green arrows) forward along the reinforced trajectory.}
    \label{fig:scenario}
\end{figure}

We consider a pure network-layer (L3) model of $N = 200$ airborne SRv6 routers $\mathcal{V} = \{0, 1, \dots, N-1\}$ moving inside a 3D operational volume $\mathcal{A} = [0, 1500]\text{ m} \times [0, 1500]\text{ m} \times [0, 300]\text{ m}$ (Figure~\ref{fig:scenario}), capturing rapid 3D aerial mobility and link transitions characteristic of emergency UAV deployments~\cite{chen2024risuav}. Each UAV $u \in \mathcal{V}$ follows a continuous, piecewise-linear 3D random waypoint trajectory $\mathbf{p}_u(t) \in \mathcal{A}$ under nominal speed profiles $v_{\text{nom}} \in \{15, 25, 35, 45\}$~m/s (sampling per-leg speeds uniformly from $[10, 20]$, $[20, 30]$, $[30, 40]$, and $[35, 50]$~m/s, respectively).

Connectivity between operational UAVs $(u,v)$ is governed by a contact range $R=220$~m:
\begin{equation}
    \mathcal{E}(t) = \left\{ (u, v) \in \mathcal{V} \times \mathcal{V} \;\middle|\; u \neq v,\; \text{up}_u(t) = 1,\; \text{up}_v(t) = 1,\; \|\mathbf{p}_u(t) - \mathbf{p}_v(t)\|_2 \le R \right\},
    \label{eq:topology}
\end{equation}
where $\text{up}_u(t) \in \{0, 1\}$ denotes operational state. Because flight legs are piecewise-linear, exact link up/down boundary times at $\|\mathbf{p}_u(t) - \mathbf{p}_v(t)\|_2 = R$ are solved analytically via quadratic root finding for $O(1)$ L3 link-state queries. Each directed link provides a $10$~Mbps rate, $20~\mu\text{s}$ processing delay, distance-dependent propagation delay, and a $128$-packet FIFO queue. Link availability is checked at both enqueue and dequeue: failure at either check invokes Stage~4 local repair while the packet remains at the sender, whereas a break after dequeue causes an in-transit loss. In practical airborne wireless interfaces, this sender-side check corresponds to hardware transmit-failure callbacks (e.g., missing link-layer acknowledgment or fast BFD/beacon loss notification) while the pre-transmission header snapshot $\sigma$ is retained in the transmit ring buffer.

\subsection{Programmable SRH Path-State Representation and Byte-Accounting Model}
\label{sec:srh_wire}
\begin{figure}[H]
    \centering
    \includegraphics[width=0.98\textwidth]{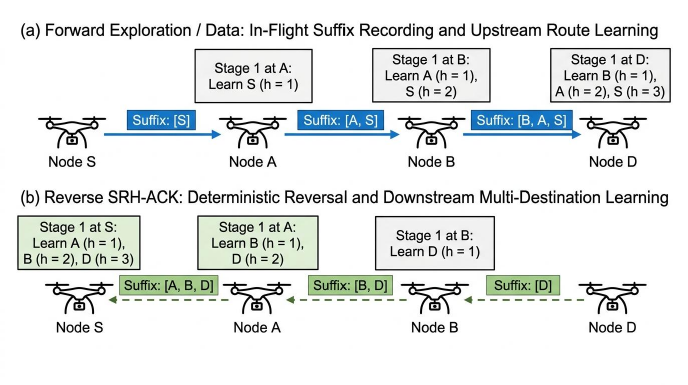}
    \caption{\textbf{INC-SRv6 in-network SRH and \texttt{SL} pointer dual-zone state mechanism across a three-hop path.} (\textbf{a})~Forward \texttt{EXPLORE}/\texttt{DATA} traversal ($S \to A \to B \to D$): relays decrement \texttt{SL}, write the selected next-hop SID into \texttt{segment\_list[SL]}, compact unused slots upon direct delivery to $D$ (see Table~\ref{tab:srh_example} for slot indices), and execute Stage~1 upstream route learning from the accumulated suffix ($[S] \to [A, S] \to [B, A, S]$ at hop distances $h=1,2,3$). (\textbf{b})~Reverse \texttt{SRH-ACK} traversal ($D \to B \to A \to S$): destination $D$ reverses the compacted array once, and each returning hop decrements \texttt{SL} deterministically without routing lookups while extracting multi-destination downstream routes from the suffix.}
    \label{fig:mechanism}
\end{figure}

INC-SRv6 operates within a trusted, homogeneous airborne SRv6 domain consistent with RFC~8754~\cite{rfc8754}, where all participating UAV routers implement the domain-local header conventions defined below. Every INC-SRv6 packet carries an outer IPv6 header ($40$~bytes) followed by a Segment Routing Header (\texttt{SRH}, Routing Type~4, $8$-byte fixed header plus populated SID entries), as illustrated in Figure~\ref{fig:mechanism}. Throughout forwarding, the \texttt{Segments Left} (\texttt{SL}) pointer satisfies the RFC~8754 invariant~\cite{rfc8754}:
\begin{equation}
    \texttt{IPv6.DA} = \texttt{SRH.segment\_list}[\texttt{SL}], \qquad \texttt{SRH.segment\_list}[0] = \text{Target SID}.
    \label{eq:sl_invariant}
\end{equation}
To record paths up to $H_{\max}=16$ hops, a router allocates a 17-slot logical segment array. A forward packet originated by source $S$ toward target $D$ places $D$ at index~$0$, $S$ at index~$16$ ($\texttt{LastEntry}=16$), and sets $\texttt{SL}=16$. As the packet traverses the network, the suffix $[\texttt{SL}..16]$ records visited routers in order, while unpopulated slots $[1..\texttt{SL}-1]$ provide local capacity for subsequent hops. At each transit relay, the router decrements $\texttt{SL}$ and writes the selected next-hop SID into $\texttt{segment\_list}[\texttt{SL}]$; when the selected next hop is the final target $D$, unused intermediate slots are compacted out and $\texttt{SL}$ is set to~$0$. Packet mode (\texttt{EXPLORE}, \texttt{ACK}, or \texttt{DATA}) and remaining split budget ($b \in \{0, 1, 2\}$) occupy domain-local bits of the 8-bit SRH \texttt{Flags} field, while the generation identifier (tracked as a monotonic 32-bit integer in the simulator) maps on the wire via modulo-$2^{16}$ reduction into the 16-bit \texttt{SRH.Tag} field ($65,536$ active generations far exceeding in-flight lifetime).

On the wire, only populated SID entries ($N_{\text{pop}}$) are serialized and charged bandwidth within the airborne SR domain:
\begin{equation}
    \texttt{LastEntry}_{\text{wire}}=N_{\text{pop}}-1,\qquad
    \texttt{SL}_{\text{wire}}=\begin{cases}1,&\text{active hop differs from target},\\0,&\text{final hop},\end{cases}
    \label{eq:wire_mapping}
\end{equation}
which maps back for forward packets at the receiving router to the logical pointer $\texttt{SL}=18-N_{\text{pop}}$ when $\texttt{SL}_{\text{wire}}=1$ (intermediate relay hop) and $\texttt{SL}=0$ when $\texttt{SL}_{\text{wire}}=0$ (final destination hop). A reverse \texttt{ACK} reverses the compacted segment array once at $D$ and decrements $\texttt{SL}$ in place along the return path. The simulator maintains the logical SRH state transitions and charges exact serialized wire bytes and transmission times according to
\begin{equation}
    B_{\text{wire}}=40+8+B_{\text{stride}}\cdot N_{\text{pop}}+B_{\text{payload}},
    \label{eq:wire_bytes}
\end{equation}
with $B_{\text{payload}}=512$~bytes for \texttt{DATA} and $0$~bytes for \texttt{EXPLORE}/\texttt{ACK}. Standard \texttt{INC-SRv6} uses 128-bit SIDs ($B_{\text{stride}}=16$~bytes), whereas \texttt{INC-SRv6-uSID} models a 4-byte compact locator stride ($B_{\text{stride}}=4$~bytes, abstracting 128-bit carrier boundary padding to isolate per-segment identifier width)~\cite{tulumello2023usid,rfc9602,rfc9800}. Non-SRv6 baseline data packets carry the $40$-byte IPv6 header plus $512$-byte payload ($552$~bytes). Table~\ref{tab:srh_example} traces the logical and wire SRH states across a three-hop traversal $S\to A\to B\to D$ and its returning acknowledgment $D\to B\to A\to S$.

\begin{table}[H]
\caption{Forward exploration and reverse acknowledgment header states. Byte columns give 128-bit / 4-byte-stride sizes with zero payload.}
\label{tab:srh_example}
\centering\footnotesize
\setlength{\tabcolsep}{3pt}
\begin{tabularx}{\linewidth}{@{}lccLcc@{}}
\toprule
\textbf{Hop} & \thead{\textbf{Logical}\\\textbf{SL}} & \thead{\textbf{Wire}\\\textbf{SL}} & \textbf{Outbound segment list} & \thead{\textbf{Split}\\\textbf{budget}} & \thead{\textbf{Bytes}\\\textbf{16B / 4B}} \\
\midrule
$S\to A$ & 15 & 1 & $[D,\mathbf{A},S]$ & 1 & 96 / 60 \\
$A\to B$ & 14 & 1 & $[D,\mathbf{B},A,S]$ & 0 & 112 / 64 \\
$B\to D$ & 0 & 0 & $[\mathbf{D},B,A,S]$ & 0 & 112 / 64 \\
\midrule
$D\to B$ (ACK) & 2 & 2 & $[S,A,\mathbf{B},D]$ & 0 & 112 / 64 \\
$B\to A$ (ACK) & 1 & 1 & $[S,\mathbf{A},B,D]$ & 0 & 112 / 64 \\
$A\to S$ (ACK) & 0 & 0 & $[\mathbf{S},A,B,D]$ & 0 & 112 / 64 \\
\bottomrule
\end{tabularx}
\end{table}

\section{Proposed INC-SRv6 Four-Stage Packet and \texorpdfstring{\texttt{SL}}{SL} Pointer Workflow}
\label{sec:workflow}

\subsection{Overview of the Four-Stage Processing Pipeline}
Each INC-SRv6 router $u \in \mathcal{V}$ maintains a local neighbor set $\mathcal{N}_u(t)$ (discovered via periodic link-local multicast \texttt{HELLO} beacons every $1.0$~s with a $3.2$~s expiration timeout evaluated on maintenance ticks) and a local normalized routing weight table $w_u(d, v) \in [0, 1]$ for each known destination $d \neq u$ and active neighbor $v \in \mathcal{N}_u(t)$, satisfying $\sum_{v \in \mathcal{N}_u(t)} w_u(d, v) = 1$. In addition, router $u$ tracks the learned hop distance $h_u(d, v)$, last reinforcement timestamp $t_{\text{last}}(u, d)$, and local repair status $\texttt{repair}_u(d)$ per destination.

Algorithm~\ref{alg:inc_srv6} gives normal packet processing; Algorithm~\ref{alg:failover} handles failed egress attempts. Stage~1 runs on arrival, Stage~2 selects forward next hops, Stage~3 writes the selected hop or advances an ACK, and Stage~4 restores and repairs a failed forward transmission. The predicate $\operatorname{confident}(d,\mathcal C)$ requires $d\notin\texttt{repair}_u$ and $\max_{v\in\mathcal C}w_u(d,v)>\theta$.

\begin{algorithmblock}{Packet processing at router $u$}
\label{alg:inc_srv6}
\begin{algorithmic}[1]
\Require Packet $p$, target $d$, mode $m$, and local routing state
\If{$p$ arrives from a neighbor}
    \State Reconcile neighbors, age accessed rows, and learn suffix (Stage~1)
    \State Release non-expired pending DATA whose routes are confident
\EndIf
\If{$u=d$}
    \If{$m=\texttt{DATA}$} \State Deliver payload
    \ElsIf{$m=\texttt{EXPLORE}$ and first arrival of this generation}
        \State Reverse path; initialize ACK at $\texttt{SL}=L-1$
        \State Enqueue ACK to its active hop with no rollback snapshot
    \EndIf
    \State \Return
\EndIf
\If{$m=\texttt{ACK}$}
    \State Decrement SL; enqueue to the indexed hop; \Return \Comment{Stage 3}
\EndIf
\If{$m=\texttt{EXPLORE}$ and generation already forwarded}
    \State \Return
\EndIf
\State If EXPLORE, mark generation forwarded; compute candidates $\mathcal C$ (Stage~2)
\If{$\mathcal C=\varnothing$}
    \State Buffer source DATA; drop transit packet; \Return
\EndIf
\If{$m=\texttt{DATA}$}
    \If{$u$ is source and route is unconfident}
        \State Buffer DATA; request cooldown-gated exploration; \Return
    \EndIf
    \State Select best candidate; refresh confident-route activity
    \State If relay is unconfident, request exploration (250~ms cooldown)
\ElsIf{route is confident and $u$ is a transit relay}
    \State Select best candidate without consuming split budget
\Else
    \State Rank by probe count, rounded weight, and seeded random tie-break
    \State Select two candidates if budget permits, otherwise one
    \State Increment selected probe counts; clone and reduce budget if splitting
\EndIf
\ForAll{selected packet/next-hop pairs $(q,v)$}
    \State If $\texttt{HopLimit} \le 0$ or ($v \neq d$ and $\texttt{SL} \le 1$), drop $q$; continue
    \State Save pre-hop snapshot $\sigma$; advance header toward $v$ (Stage~3)
    \State Enqueue $(q,v,\sigma)$; check link at enqueue and dequeue
\EndFor
\end{algorithmic}
\end{algorithmblock}

\begin{algorithmblock}{Egress-failure callback (Stage 4)}
\label{alg:failover}
\begin{algorithmic}[1]
\Require Failed next hop $v$, packet $p$, target $d$, pre-hop snapshot $\sigma$
\State Remove $v$; age and renormalize weights; mark target $d$ for repair
\If{$p$ is ACK or $\sigma$ is absent}
    \State Drop $p$; \Return
\EndIf
\State Restore the complete snapshot $\sigma$
\State Recompute loop-free candidates $\mathcal C$ from the restored history
\If{$p$ is EXPLORE}
    \State Set split budget to zero and retain generation identity
    \If{$\mathcal C=\varnothing$} \State Drop $p$; \Return \EndIf
    \State Select best candidate $v'$; if Stage~3 hop/$\texttt{SL}$ bound fails, drop $p$ and \Return
    \State Save snapshot, advance toward $v'$ (Stage~3), and enqueue checked transmission
\ElsIf{$\mathcal C=\varnothing$}
    \State Buffer source DATA; drop relay DATA
\ElsIf{$u$ is the DATA source}
    \State Buffer DATA; originate exploration with cooldown bypass
\Else
    \State Originate exploration with cooldown bypass; select best candidate $v'$
    \State If Stage~3 hop/$\texttt{SL}$ bound fails, drop $p$ and \Return
    \State Save snapshot, advance relay DATA toward $v'$ (Stage~3), and enqueue transmission
\EndIf
\end{algorithmic}
\end{algorithmblock}

\subsection{Stage 1: Ingress SRH Suffix Learning (Multi-Hop Path Extraction)}
\label{sec:stage1}
Before looking up or updating destination $d$, router $u$ reconciles weight row $w_u(d, \cdot)$ with its active neighbor set $\mathcal{N}_u(t)$ and applies lazy exponential aging toward the uniform prior $1/m$ ($m = |\mathcal{N}_u(t)|$) over idle interval $\Delta t = t - t_{\text{last}}(u, d)$:
\begin{equation}
    w_u(d, v) \leftarrow \frac{1}{m} + \left( w_u(d, v) - \frac{1}{m} \right) \exp\!\left( -\frac{\Delta t}{\tau} \right), \qquad \forall v \in \mathcal{N}_u(t),
    \label{eq:aging}
\end{equation}
with $\tau = 8.0$~s, followed by normalization $\sum_{v \in \mathcal{N}_u(t)} w_u(d, v) = 1$. If $u$ has forwarded \texttt{DATA} toward $d$ on a confident route within the preceding $250$~ms activity window, aging on $d$ is suspended and $t_{\text{last}}(u, d)$ is refreshed. When $k_{\text{new}}$ new neighbors join $\mathcal{N}_u(t)$ while $u$ holds a confident non-repair route to $d$ ($d \notin \texttt{repair}_u$ and $w_{\max} = \max_v w_u(d, v) > \theta$), their combined prior share is capped at $\min\!\left(k_{\text{new}}/m,\, 0.08,\, \max(0, 1 - (\theta+0.02)/w_{\max})\right)$ to preserve the $\theta + 0.02$ confidence margin.

For each arriving suffix index $k>\texttt{SL}$, let $x_k=\texttt{segment\_list}[k]$ denote the learned destination, $v_{\text{prev}}=\texttt{segment\_list}[\texttt{SL}+1]$ denote the immediate upstream sender, and $h=k-\texttt{SL}$ denote the exact hop distance from $u$ to $x_k$ along that suffix. Forward \texttt{EXPLORE} and \texttt{DATA} suffixes teach routes toward upstream origins and relays, whereas returning \texttt{ACK} suffixes teach validated routes toward the target destination and downstream relays. Each tuple $(\text{generation},\text{mode},x_k)$ is learned at most once per router, and an observation is ignored if $u$ already holds a confident, non-repair route to $x_k$ with a strictly smaller recorded hop distance than~$h$.

When $(x_k, h)$ passes deduplication and short-route protection, router $u$ reinforces neighbor $v_{\text{prev}}$ using a distance-attenuated step size $\alpha(h)$:
\begin{equation}
    \alpha(h) = \min\!\left( \alpha_0,\; 0.52 + \frac{\max(0,\, \alpha_0 - 0.52)}{h} \right), \qquad \alpha_0 = 0.85,
    \label{eq:alpha}
\end{equation}
and updates the routing weights via convex combination (clearing $x_k$ from $\texttt{repair}_u$ once $\max_v w_u^+(x_k, v) > \theta$):
\begin{equation}
    w_u^+(x_k, v) = (1 - \alpha(h))\, w_u(x_k, v) + \alpha(h)\, \mathbb{I}(v = v_{\text{prev}}), \qquad \forall v \in \mathcal{N}_u(t).
    \label{eq:update}
\end{equation}
Because $\alpha(h) \ge 0.52 + 0.33/16 = 0.5406 > \theta = 0.51$ for all $1\le h\le16$, a single validated observation elevates $v_{\text{prev}}$ strictly above $\theta$ in one pass. For example, from a four-neighbor uniform prior ($w=0.25$), a two-hop \texttt{ACK} ($h=2$) applies $\alpha(2)=0.685$ and raises the reinforced neighbor's weight to $w^+=0.764$ in one RTT (Figure~\ref{fig:stage2}b).

\subsection{Stage 2: Confidence-Gated Next-Hop Selection and Bounded Cloning}
\label{sec:stage2}
\begin{figure}[H]
    \centering
    \includegraphics[width=0.98\textwidth]{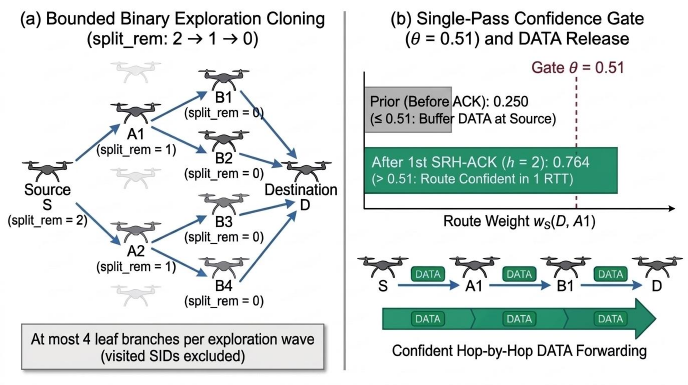}
    \caption{\textbf{Stage 2: Bounded Exploration Cloning and Confidence-Gated Next-Hop Selection.} (\textbf{a})~Least-probed bounded binary exploration cloning tree ($\texttt{split\_rem}: 2\to1\to0$), restricting each unconfident exploration wave to at most $4$ leaf branches while excluding visited suffix nodes. (\textbf{b})~Single-pass confidence gate ($\theta=0.51$) crossed in one RTT (illustrated for a representative $h=2$ returning \texttt{SRH-ACK} with $\alpha(2)=0.685$, elevating $w_S(D,A_1)$ from $0.250$ to $0.764$; a three-hop $h=3$ return similarly yields $\alpha(3)=0.630$ and $w^+=0.723 > \theta$) to release buffered \texttt{DATA}.}
    \label{fig:stage2}
\end{figure}

Figure~\ref{fig:stage2} illustrates bounded exploration cloning and the single-pass confidence gate. For any transit packet ($\texttt{SL} > 0$, target $d \neq u$), Stage~2 forms the loop-free candidate set by excluding all routers already recorded in the traversed suffix $V_{\text{visited}} = \texttt{SRH.segment\_list}[\texttt{SL}..\texttt{LastEntry}]$:
\begin{equation}
    \mathcal{C}_u(d) = \left\{ v \in \mathcal{N}_u(t) \;\middle|\; \text{SID}(v) \notin V_{\text{visited}} \right\}.
    \label{eq:candidates}
\end{equation}
Router $u$ is confident for $d$ when $d\notin\texttt{repair}_u$ and $\max_{v\in\mathcal{C}_u(d)} w_u(d,v) > \theta = 0.51$. For \texttt{DATA} packets, an unconfident source enqueues the payload in a per-destination FIFO buffer (bounded at $64$ packets with $1.0$-s expiration) and triggers an \texttt{EXPLORE} wave ($50$-ms source cooldown, retried every $250$~ms on maintenance ticks); any Stage~1 update that renders $d$ confident immediately releases all buffered \texttt{DATA}. Transit relays forward \texttt{DATA} greedily along the highest-weight candidate in $\mathcal{C}_u(d)$ and trigger a background \texttt{EXPLORE} probe ($250$-ms relay cooldown) if locally unconfident.

For \texttt{EXPLORE} packets, each generation is forwarded at most once per relay. A confident transit relay steers the probe along its single best candidate without decrementing the split budget~$b$. At the source or an unconfident relay, Stage~2 ranks $\mathcal{C}_u(d)$ by ascending probe count $c_{\text{probe}}(d,v)$, descending weight, and a seeded tie-breaker. When $b>0$ and $|\mathcal{C}_u(d)| \ge 2$, the router clones the packet across the top two candidates and decrements the child budget:
\begin{equation}
    b_{\text{child}}=b-1\quad\text{when splitting},\qquad
    c_{\text{probe}}(d,v)\leftarrow c_{\text{probe}}(d,v)+1.
    \label{eq:cloning}
\end{equation}
With $b_{\max}=2$ at the source, binary splitting occurs at most twice along any path, bounding each discovery wave to at most $2^2=4$ leaf branches (Figure~\ref{fig:stage2}a).

\subsection{Stage 3: \texorpdfstring{\texttt{SL}}{SL} Pointer Advancement and Deterministic \texorpdfstring{\texttt{SRH-ACK}}{SRH-ACK} Reversal}
\label{sec:stage3}
For each outbound forward branch $(q, v_{\text{next}})$ with positive $\texttt{HopLimit}$, Stage~3 saves a pre-transmission snapshot $\sigma = (\texttt{segment\_list}, \texttt{SL}, \texttt{IPv6.DA}, \texttt{HopLimit})$. If $v_{\text{next}} = d$, Stage~3 compacts unused slots into $\texttt{segment\_list} \leftarrow [d] \cup \texttt{suffix}$ and sets $\texttt{SL} \leftarrow 0$. Otherwise, if $v_{\text{next}} \neq d$ and $\texttt{SL} \le 1$ (meaning the branch has exhausted the $H_{\max}=16$ intermediate segment capacity without reaching $d$), or if $\texttt{HopLimit} \le 0$, the router drops the branch to enforce the strict path-depth bound. For a valid intermediate hop ($v_{\text{next}} \neq d$ and $\texttt{SL} > 1$), Stage~3 advances the pointer, decrements $\texttt{HopLimit}$, and writes the next-hop SID:
\begin{equation}
    \begin{aligned}
    \texttt{SL}&\leftarrow\texttt{SL}-1,\\
    \texttt{segment\_list}[\texttt{SL}]&\leftarrow\operatorname{SID}(v_{\text{next}}),\\
    \texttt{IPv6.DA}&\leftarrow\texttt{segment\_list}[\texttt{SL}].
    \end{aligned}
    \label{eq:forward_sl}
\end{equation}
When the first \texttt{EXPLORE} branch of a generation reaches destination $D$ ($\texttt{SL}=0$) with compacted segment array $[D,r_1,\ldots,r_{L-1},S]$, $D$ reverses the array once into
\begin{equation}
    \texttt{segment\_list}_{\text{ACK}}=[S,r_{L-1},\ldots,r_1,D],
    \label{eq:ack_reverse}
\end{equation}
sets $\texttt{SL}=L-1$ and $\texttt{IPv6.DA}=r_1$, and transmits the unicast \texttt{ACK}. Each returning hop executes Stage~1 suffix learning and decrements the pointer in place:
\begin{equation}
    \texttt{SL}\leftarrow\texttt{SL}-1,\qquad
    \texttt{IPv6.DA}\leftarrow\texttt{segment\_list}_{\text{ACK}}[\texttt{SL}],
    \label{eq:ack_step}
\end{equation}
until reaching source $S$ at $\texttt{SL}=0$ (Table~\ref{tab:srh_example}).

\subsection{Stage 4: Local Link-Down Failover and Table Aging}
\label{sec:stage4}
\begin{figure}[H]
    \centering
    \includegraphics[width=0.98\textwidth]{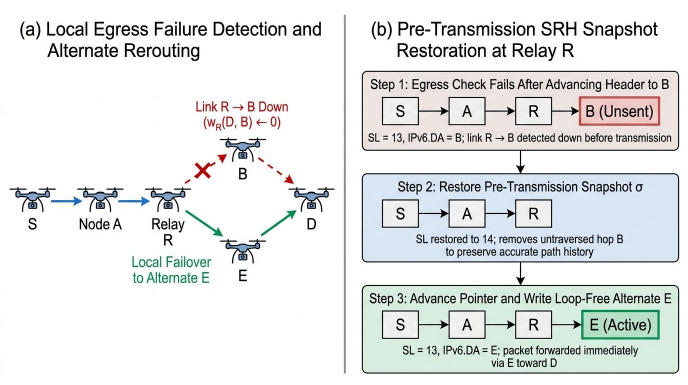}
    \caption{\textbf{Stage 4: Local Link-Down Failover and Pre-Transmission SRH Snapshot Restoration.} (\textbf{a})~Relay $R$ detects a broken egress link to primary next hop $B$, removes $B$ from $\mathcal{N}_R(t)$, renormalizes surviving weights, and reroutes immediately to loop-free alternate $E$ while triggering a background probe (with idle entries aging exponentially at $\tau=8.0$~s). (\textbf{b})~Three-step header rollback and rewrite along the logical traversal chain: the failed attempt at $R \to B$ had decremented $\texttt{SL}$ from $14$ to $13$ (Step~1); restoring snapshot $\sigma$ resets $\texttt{SL}=14$ and expunges unsent hop $B$ (Step~2) before rewriting alternate $E$ at $\texttt{SL}=13$ (Step~3).}
    \label{fig:stage4}
\end{figure}

When the enqueue or dequeue check at router $u$ detects that next hop $v_{\text{fail}}$ is unreachable, Stage~4 executes immediate local failover (Algorithm~\ref{alg:failover} and Figure~\ref{fig:stage4}). First, $u$ removes $v_{\text{fail}}$ from $\mathcal{N}_u(t)$, marks $d \in \texttt{repair}_u$, and renormalizes remaining neighbor weights:
\begin{equation}
    w_u(d,v)\leftarrow
    \begin{cases}
    w_u(d,v)/\sum_{z\in\mathcal N_u(t)}w_u(d,z),&\text{positive sum},\\
    1/|\mathcal N_u(t)|,&\text{zero sum and nonempty set}.
    \end{cases}
    \label{eq:failover_norm}
\end{equation}
Second, for a forward packet carrying snapshot $\sigma$, Stage~4 restores $\sigma$ (\texttt{segment\_list}, \texttt{SL}, \texttt{IPv6.DA}, and \texttt{HopLimit}) to its exact pre-hop arrival state. Because Stage~3 had already decremented $\texttt{SL}$ and written $\operatorname{SID}(v_{\text{fail}})$ into the header prior to the link check, restoring $\sigma$ prevents the untraversed node $v_{\text{fail}}$ from polluting the suffix or consuming an extra segment slot. Third, $u$ recomputes the loop-free candidates $\mathcal{C}_u(d)$. A failed \texttt{EXPLORE} branch sets its split budget to zero and advances to the best surviving candidate; a failed \texttt{DATA} packet at a transit relay triggers a cooldown-bypassed \texttt{EXPLORE} probe and immediately re-advances the restored packet toward the highest-weight surviving candidate in $\mathcal{C}_u(d)$ (e.g., rerouting from failed hop $B$ to alternate $E$ in Figure~\ref{fig:stage4}).

\section{Performance Evaluation}
\label{sec:evaluation}

\subsection{Pure-L3 Simulation Setup, Baselines, and Timing Definitions}
We implement the pure-L3 discrete-event simulator, the proposed INC-SRv6 mechanism, all baseline protocols, and the evaluation pipeline in Python~3.10 (verified by 64 deterministic unit tests; open-source repository details are provided in the Data Availability Statement). Table~\ref{tab:params} summarizes the physical, network-layer, and protocol parameters used across all experiments. Each simulation run spans $50.0$~s ($40.0$~s of active traffic generation followed by a $10.0$~s drain window) across $N=200$ UAVs. Traffic consists of $40$ concurrent constant-bit-rate (CBR) flows between randomly sampled distinct source-destination pairs, each generating $512$-byte UDP payloads at $8$~packets/s ($32.768$~kbps payload rate per flow, $125$~ms inter-packet interval), with flow start times staggered uniformly in $[2.0, 4.0]$~s. All periodic control messages (\texttt{HELLO}, \texttt{LSA}, \texttt{TC}, and \texttt{RREQ}) are transmitted via link-local L3 multicast, charging one serialization event at the sender's multicast FIFO and delivering independent copies to all currently connected L3 neighbors. Note that this collision-free pure-L3 multicast abstraction is intentionally conservative in favor of broadcast-heavy baselines (\texttt{AODV-L3}, \texttt{OLSR-L3}, \texttt{OSPF-SRv6}, and \texttt{AntHocNet-L3}): under realistic CSMA/CA wireless contention, network-wide \texttt{RREQ} and \texttt{LSA}/\texttt{TC} broadcasts suffer severe broadcast-storm collisions and contention backoffs, whereas INC-SRv6 relies on bounded unicast exploration ($\le 4$ leaf branches) and passive suffix extraction.

\begin{table}[H]
\caption{Pure-L3 3D UAV network simulation parameters and INC-SRv6 configuration.}
\label{tab:params}
\centering
\footnotesize
\setlength{\tabcolsep}{4.5pt}
\begin{tabularx}{\linewidth}{llL}
\toprule
\textbf{Category} & \textbf{Parameter} & \textbf{Value / Engineering Description} \\
\midrule
\multirow{5}{*}{3D UAV Topology} & Number of UAV routers ($N$) & $200$ airborne SRv6 routers \\
 & Operational airspace ($\mathcal{A}$) & $1500\text{ m} \times 1500\text{ m} \times 300\text{ m}$ (3D Random Waypoint) \\
 & L3 communication range ($R$) & $220\text{ m}$ (mean geometric node degree $\approx 14.8$--$15.2$) \\
 & Simulation horizon & $40.0\text{ s}$ active traffic + $10.0\text{ s}$ drain ($50.0\text{ s}$ total) \\
 & Paired random seeds & $10$ paired seeds per scenario ($\texttt{seed} \in \{1001, \dots, 1010\}$) \\
\midrule
\multirow{5}{*}{Pure-L3 Forwarding} & Directed link capacity ($C_{\text{link}}$) & $10\text{ Mbps}$ point-to-point L3 pipe (zero MAC retry) \\
 & Per-hop SRH processing ($d_{\text{proc}}$) & $20~\mu\text{s}$ per forwarded packet \\
 & Interface FIFO queue ($Q_{\max}$) & $128$ packets per directed interface \\
 & Offered CBR workload & $40$ concurrent flows, $8\text{ pkts/s/flow}$, $512\text{ B}$ payload \\
 & Modeled byte accounting & $40\text{ B}$ IPv6 + $8\text{ B}$ SRH base + $B_{\text{stride}} \times N_{\text{pop}}$ ($16\text{ B}$ SID / $4\text{ B}$ uSID model) \\
\midrule
\multirow{6}{*}{INC-SRv6 Parameters} & Confidence gate ($\theta$) & $0.51$ (strict majority confidence at source) \\
 & Reinforcement step ($\alpha_0, \alpha(h)$) & $\alpha_0 = 0.85$, $\alpha(h) = \min(\alpha_0, 0.52 + 0.33/h)$ (Equation~\eqref{eq:alpha}) \\
 & Bounded cloning budget ($b_{\max}$) & $\texttt{split\_rem} = 2$ ($\le 4$ leaf branches per exploration wave) \\
 & Maximum SRH path depth ($H_{\max}$) & $16$ hops ($17$ logical segment slots in \texttt{SRH.segment\_list}) \\
 & Table aging \& active refresh & $\tau = 8.0\text{ s}$ lazy exponential decay; $250\text{ ms}$ active \texttt{DATA} window \\
 & Source pending buffer & Max $64\text{ pkts/destination}$, $1.0\text{ s}$ wait expiration \\
\bottomrule
\end{tabularx}

\end{table}

Five pure-L3 engineering baselines share the common L3 forwarding engine (establishing controlled algorithmic reference baselines within the shared L3 discrete-event engine): (i)~\texttt{OSPF-SRv6}~\cite{rfc5340} disseminates neighbor LSAs every $2.0$~s (batched over a $10$-ms window), computes shortest-path source SRH segment lists from its local LSDB, and performs local LFA/TI-LFA style alternate repair upon egress link loss using a cached reverse shortest-path tree excluding the repair node; (ii)~\texttt{OLSR-L3} transmits $1.0$-s \texttt{HELLO} beacons and $2.0$-s MPR-relayed \texttt{TC} advertisements to maintain hop-by-hop shortest paths; (iii)~\texttt{AODV-L3} discovers scalar next-hop routes via flooded \texttt{RREQ} and unicast \texttt{RREP} messages ($8.0$-s active route lifetime, $500$-ms per-destination request cooldown) and propagates \texttt{RERR} notifications to upstream precursors upon link failure to trigger source rediscovery; (iv)~\texttt{Q-Routing-L3} updates per-destination/neighbor delivery delay estimates via charged one-hop feedback packets ($\eta=0.3$, $\epsilon=0.1$ exploration); and (v)~\texttt{AntHocNet-L3} combines reactive forward/backward ants with $2.0$-s proactive path probing, periodic pheromone evaporation ($0.9$ factor), and neighbor pheromone diffusion.

To isolate the contribution of each design element in our workflow, four mechanism ablations modify exactly one component of full \texttt{INC-SRv6}: \texttt{INC-SRv6-OriginOnly} restricts Stage~1 suffix learning to the packet's origin node ($k = \texttt{LastEntry}$) rather than the full traversed suffix; \texttt{INC-SRv6-NoClone} disables Stage~2 binary exploration cloning ($b_{\max}=0$, single-branch probing); \texttt{INC-SRv6-SmallStep} replaces the single-pass reinforcement rule in Equation~\eqref{eq:alpha} with a weak step size $\alpha(h)=0.35/h < \theta$; and \texttt{INC-SRv6-NoFastFailover} disables Stage~4 immediate neighbor invalidation, snapshot restoration, and local rerouting upon link-down events, relying instead on periodic \texttt{HELLO} expiration. In addition, \texttt{INC-SRv6-uSID} evaluates the 4-byte compact SID stride ($B_{\text{stride}}=4$~bytes).

End-to-end packet delivery ratio (PDR) is defined as unique delivered \texttt{DATA} payloads divided by total generated \texttt{DATA} payloads across the 40 flows. End-to-end delay $t_{\text{deliver}}-t_{\text{born}}$ is measured over all delivered \texttt{DATA} packets and includes source pending-buffer waiting time, interface FIFO queuing, $20~\mu\text{s}$ per-hop processing, serialization, and propagation delay. Each table entry reports the arithmetic mean across the $10$ paired seeds, where $p_{95}$ delay is computed per seed and averaged across seeds. Control overhead counts all on-wire control hop transmissions (\texttt{HELLO}, \texttt{EXPLORE}, \texttt{ACK}, \texttt{LSA}, \texttt{TC}, \texttt{RREQ}, \texttt{RREP}, \texttt{RERR}, \texttt{FANT}, \texttt{BANT}, \texttt{DIFFUSE}, and \texttt{QFEEDBACK}) over the $50.0$-s simulation horizon.

Table~\ref{tab:scenarios} defines the ten evaluation scenarios ($\texttt{S1}$--$\texttt{S10}$), covering pure mobility scaling from $15$ to $45$~m/s ($\texttt{S1}$--$\texttt{S4}$), central high-degree relay hub failures ($\texttt{S5}$--$\texttt{S6}$), contiguous spatial area blackouts ($\texttt{S7}$), repeated multi-wave relay churn ($\texttt{S8}$--$\texttt{S9}$), and combined high-speed flight with repeated $20\%$ relay churn ($\texttt{S10}$). Relay failures affect only intermediate transit nodes while keeping all flow endpoints operational. Every protocol and variant executes on the exact same precomputed 3D trajectories, flow schedules, and failure cohorts for each seed, yielding $1,100$ paired simulation runs.

\begin{table}[H]
\caption{Experimental scenario matrix ($\texttt{S1}$--$\texttt{S10}$) for the 200-node 3D UAV network campaign.}
\label{tab:scenarios}
\centering
\footnotesize
\setlength{\tabcolsep}{4pt}
\begin{tabularx}{\linewidth}{llccL}
\toprule
\textbf{Scenario ID} & \textbf{Category} & \thead{\textbf{Speed Range}\\\textbf{(m/s)}} & \thead{\textbf{Relay}\\\textbf{Outage}} & \textbf{Operational / Fault Injection Description} \\
\midrule
\texttt{S1} & Mobility & $15$ ($[10, 20]$) & $0\%$ & Normal patrol cruise speed, zero node failures \\
\texttt{S2} & Mobility & $25$ ($[20, 30]$) & $0\%$ & Fast transit cruise speed, zero node failures \\
\texttt{S3} & Mobility & $35$ ($[30, 40]$) & $0\%$ & High-speed dash maneuver, rapid link turnover \\
\texttt{S4} & Mobility & $45$ ($[35, 50]$) & $0\%$ & Extreme aerial speed, severe link churn \\
\midrule
\texttt{S5} & Hub Outage & $25$ ($[20, 30]$) & $10\%$ ($20$ relays) & Top-$20$ highest-degree central relays fail at $t=15.0$~s \\
\texttt{S6} & Hub Outage & $25$ ($[20, 30]$) & $20\%$ ($40$ relays) & Top-$40$ highest-degree central relays fail at $t=15.0$~s \\
\texttt{S7} & Spatial Blackout & $25$ ($[20, 30]$) & $20\%$ ($40$ relays) & $40$ relays nearest a seeded reference UAV position fail at $t=15.0$~s \\
\midrule
\texttt{S8} & Churn & $25$ ($[20, 30]$) & $3 \times 10\%$ & Three $4$-s relay outage waves ($20$ relays each) at $t=12, 22, 32$~s \\
\texttt{S9} & Churn & $25$ ($[20, 30]$) & $3 \times 20\%$ & Three $4$-s relay outage waves ($40$ relays each) at $t=12, 22, 32$~s \\
\texttt{S10} & Combined & $35$ ($[30, 40]$) & $3 \times 20\%$ & High-speed flight ($35$~m/s) + three $4$-s, $40$-relay outage waves \\
\bottomrule
\end{tabularx}

\end{table}

\subsection{Mobility Scaling Analysis (\texorpdfstring{$\texttt{S1}$--$\texttt{S4}$}{S1--S4})}
\begin{figure}[H]
    \centering
    \includegraphics[width=0.98\textwidth]{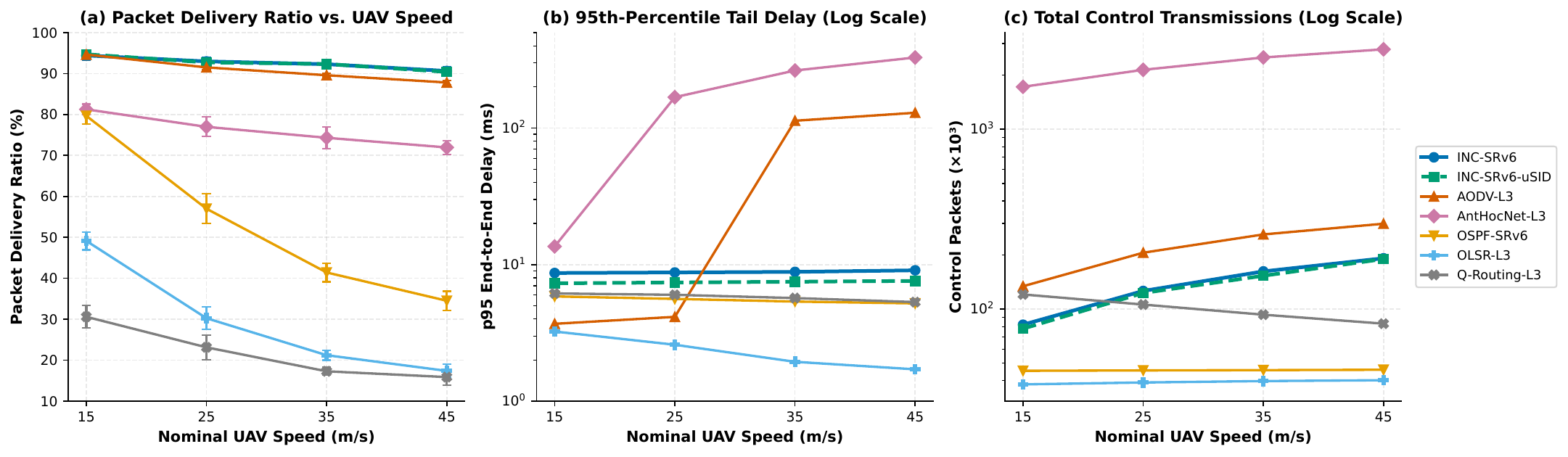}
    \caption{\textbf{Mobility scaling performance across $\texttt{S1}$ ($15$~m/s) to $\texttt{S4}$ ($45$~m/s).} (\textbf{a})~Mean packet delivery ratio (PDR, $\%$) with $95\%$ confidence intervals across 10 paired seeds. (\textbf{b})~95th-percentile ($p_{95}$) end-to-end tail delay (ms, log scale). (\textbf{c})~Total on-wire control packet transmissions ($\times 10^3$, log scale).}
    \label{fig:mobility}
\end{figure}

Figure~\ref{fig:mobility} and Table~\ref{tab:mobility_results} present the delivery ratio, mean delay, 95th-percentile ($p_{95}$) tail latency, and control overhead as nominal UAV speed scales from $15$~m/s (\texttt{S1}) to $45$~m/s (\texttt{S4}). Key performance figures are underlined in Table~\ref{tab:mobility_results} for ease of comparison. Across high-delivery regimes in Tables~\ref{tab:mobility_results} and~\ref{tab:fault_results}, arithmetic mean delay can slightly exceed $p_{95}$ delay (e.g., \texttt{INC-SRv6} in $\texttt{S3}$--$\texttt{S4}$ and \texttt{AODV-L3} in $\texttt{S6}$--$\texttt{S7}$): over $95\%$ of delivered payloads traverse established routes within a few milliseconds, whereas fewer than $5\%$ of initial or post-break packets wait hundreds of milliseconds (up to the $1.0$-s buffer cap) in the source pending queue during route discovery, pulling up the arithmetic mean without shifting the 95th percentile.

\begin{table}[H]
\caption{Mobility scaling performance comparison across $\texttt{S1}$ ($15$~m/s), $\texttt{S3}$ ($35$~m/s), and $\texttt{S4}$ ($45$~m/s) (10-seed arithmetic means; key comparison values are underlined).}
\label{tab:mobility_results}
\centering
\footnotesize
\setlength{\tabcolsep}{4pt}
\begin{tabularx}{\linewidth}{llRRRRR}
\toprule
\textbf{Scenario} & \textbf{Protocol} & \rheadcell{\textbf{PDR}\\\textbf{(\%)}} & \rheadcell{\textbf{Mean}\\\textbf{Delay (ms)}} & \rheadcell{\textbf{$p_{95}$}\\\textbf{Delay (ms)}} & \rheadcell{\textbf{Control}\\\textbf{Pkts}} & \rheadcell{\textbf{Control}\\\textbf{MB}} \\
\midrule
\multirow{7}{*}{\texttt{S1}} & \textbf{INC-SRv6} & $94.40$ & $8.11$ & $8.68$ & $81,930$ & $13.54$ \\
 & \textbf{INC-SRv6-uSID} & \underline{$\mathbf{94.70}$} & $7.47$ & \underline{$7.29$} & \underline{$77,856$} & \underline{$\mathbf{6.44}$} \\
 & AODV-L3 & \underline{$94.63$} & $4.86$ & \underline{$\mathbf{3.68}$} & \underline{$134,072$} & \underline{$8.02$} \\
 & AntHocNet-L3 & $81.25$ & $19.19$ & $13.58$ & $1,723,589$ & $159.69$ \\
 & OSPF-SRv6 & $79.67$ & $2.74$ & $5.85$ & $45,404$ & $78.61$ \\
 & OLSR-L3 & $49.14$ & $1.57$ & $3.24$ & $38,148$ & $59.50$ \\
 & Q-Routing-L3 & $30.60$ & $2.50$ & $6.15$ & $120,672$ & $7.68$ \\
\midrule
\multirow{7}{*}{\texttt{S3}} & \textbf{INC-SRv6} & \underline{$92.28$} & $9.89$ & \underline{$8.85$} & $162,155$ & $27.61$ \\
 & \textbf{INC-SRv6-uSID} & \underline{$\mathbf{92.33}$} & $\mathbf{9.17}$ & \underline{$\mathbf{7.49}$} & \underline{$153,329$} & \underline{$\mathbf{12.51}$} \\
 & AODV-L3 & \underline{$89.59$} & $14.47$ & \underline{$113.07$} & \underline{$260,248$} & $15.19$ \\
 & AntHocNet-L3 & $74.31$ & $39.74$ & $263.81$ & $2,509,042$ & $230.87$ \\
 & OSPF-SRv6 & $41.42$ & $2.24$ & $5.36$ & $45,799$ & $88.15$ \\
 & OLSR-L3 & $21.22$ & $1.00$ & $1.94$ & $39,808$ & $67.05$ \\
 & Q-Routing-L3 & $17.29$ & $2.04$ & $5.68$ & $93,126$ & $6.09$ \\
\midrule
\multirow{7}{*}{\texttt{S4}} & \textbf{INC-SRv6} & \underline{$\mathbf{90.63}$} & $11.55$ & \underline{$9.08$} & $191,806$ & $33.07$ \\
 & \textbf{INC-SRv6-uSID} & \underline{$90.39$} & $\mathbf{11.19}$ & \underline{$\mathbf{7.59}$} & \underline{$189,864$} & \underline{$\mathbf{15.49}$} \\
 & AODV-L3 & \underline{$87.81$} & $18.79$ & \underline{$129.37$} & \underline{$298,515$} & $17.37$ \\
 & AntHocNet-L3 & $71.95$ & $46.36$ & $328.47$ & $2,784,461$ & $257.34$ \\
 & OSPF-SRv6 & $34.51$ & $2.05$ & $5.20$ & $45,997$ & $91.01$ \\
 & OLSR-L3 & $17.39$ & $0.86$ & $1.71$ & $40,236$ & $69.13$ \\
 & Q-Routing-L3 & $15.88$ & $1.79$ & $5.31$ & $83,018$ & $5.51$ \\
\bottomrule
\end{tabularx}

\end{table}

The mobility scaling results in Figure~\ref{fig:mobility} and Table~\ref{tab:mobility_results} highlight two distinct operating regimes:
\begin{itemize}[leftmargin=1.5em]
    \item \textbf{Moderate Cruise Regime ($\texttt{S1}$, $15$~m/s):} Both \texttt{INC-SRv6-uSID} ($94.70\%$ PDR; $94.40\%$ for 128-bit \texttt{INC-SRv6}) and reactive \texttt{AODV-L3} ($94.63\%$ PDR) achieve near-ceiling delivery. However, whereas \texttt{AODV-L3} floods \texttt{RREQ} broadcasts across all 200 nodes upon route expiration ($134,072$ control packets, $8.02$~MB), INC-SRv6 reuses suffixes learned from intersecting flows and restricts unconfident probing to at most four leaf branches, reducing control packets by $41.9\%$ ($77,856$) and control bytes by $19.7\%$ ($6.44$~MB) under \texttt{uSID}. Because \texttt{AODV-L3} pins scalar shortest-hop routes and drops packets upon link loss rather than rerouting locally, its surviving payloads exhibit lower $p_{95}$ delay ($3.68$~ms versus $7.29$~ms for \texttt{INC-SRv6-uSID}).
    \item \textbf{High-Speed Maneuver Regime ($\texttt{S2}$--$\texttt{S4}$, $25$--$45$~m/s):} Rapid link turnover continuously breaks active paths. Under \texttt{AODV-L3}, intermediate link breaks drop in-flight packets, propagate \texttt{RERR} messages upstream, and stall sources behind discovery cooldowns until a new \texttt{RREQ}/\texttt{RREP} flood completes, degrading PDR to $87.81\%$ in $\texttt{S4}$ while inflating $p_{95}$ tail delay by over $30\times$ from $3.68$~ms ($\texttt{S1}$) to $113.07$~ms ($\texttt{S3}$) and $129.37$~ms ($\texttt{S4}$). Conversely, \texttt{INC-SRv6} executes Stage~4 snapshot rollback and local alternate forwarding at the detecting relay, sustaining $92.28\%$ ($\texttt{S3}$) and $90.63\%$ ($\texttt{S4}$) PDR while bounding $p_{95}$ delay tightly at $8.85$--$9.08$~ms ($7.49$--$7.59$~ms for \texttt{uSID}), achieving a $>92\%$ tail-latency reduction alongside $36.4\%$--$41.1\%$ fewer control packets than \texttt{AODV-L3}.
    \item \textbf{Proactive, Learning, and Swarm Baselines:} Proactive \texttt{OSPF-SRv6} and \texttt{OLSR-L3} fall from $79.67\%$ and $49.14\%$ PDR in $\texttt{S1}$ to $34.51\%$ and $17.39\%$ in $\texttt{S4}$ while consuming $59.50$--$91.01$~MB of control traffic, as 2-s topology advertisements lag behind sub-second link changes. \texttt{Q-Routing-L3} delivers only $15.88\%$--$30.60\%$ because one-hop feedback propagates too slowly across multi-hop trajectories, whereas \texttt{AntHocNet-L3} achieves $71.95\%$--$81.25\%$ PDR at the cost of $13.58$--$328.47$~ms $p_{95}$ delay ($168.19$--$328.47$~ms across $25$--$45$~m/s) and $1.72$--$2.78$~million control packets ($159.69$--$257.34$~MB).
\end{itemize}

\subsection{Resilience Under Relay Hub Faults, Spatial Blackouts, and Repeated Churn (\texorpdfstring{$\texttt{S5}$--$\texttt{S10}$}{S5--S10})}
\begin{figure}[H]
    \centering
    \includegraphics[width=0.98\textwidth]{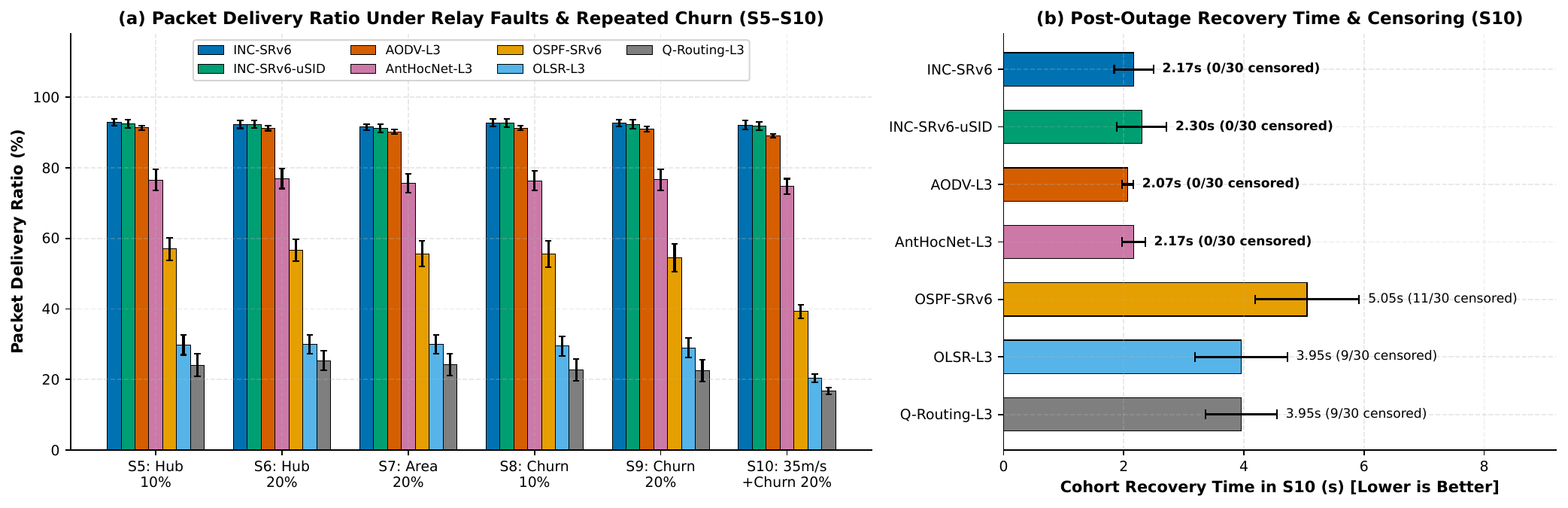}
    \caption{\textbf{Fault resilience and post-outage recovery across $\texttt{S5}$--$\texttt{S10}$.} (\textbf{a})~Mean packet delivery ratio (PDR, $\%$) with $95\%$ confidence intervals under single hub faults ($\texttt{S5}$--$\texttt{S6}$), spatial area blackout ($\texttt{S7}$), repeated relay churn ($\texttt{S8}$--$\texttt{S9}$), and high-speed churn ($\texttt{S10}$). (\textbf{b})~Post-outage cohort recovery time (s) and fractions of events right-censored at the observation limit across the 30 outage events in $\texttt{S10}$.}
    \label{fig:faults}
\end{figure}

Figure~\ref{fig:faults} and Table~\ref{tab:fault_results} evaluate protocol resilience across the relay failure scenarios ($\texttt{S5}$--$\texttt{S10}$). For each outage event, cohort recovery time measures the time required for two consecutive $1$-s payload generation cohorts over a $10$-s post-fault start window to recover at least $90\%$ of the pre-fault $4$-s baseline PDR (minimum recovery window $2.00$~s, observed completion up to $11.0$~s after single outages or $8.0$~s for the final $t=32$~s wave before the $40$-s generation cutoff; reported values are arithmetic means over uncensored events, and the censoring count denotes non-recovery within that window).

\begin{table}[H]
\caption{Fault resilience and post-outage recovery comparison across $20\%$ hub fault ($\texttt{S6}$), $20\%$ spatial blackout ($\texttt{S7}$), and $35$~m/s high-speed repeated $20\%$ relay churn ($\texttt{S10}$) (10-seed arithmetic means; key comparison values are underlined).}
\label{tab:fault_results}
\centering
\scriptsize
\setlength{\tabcolsep}{3.5pt}
\begin{tabularx}{\linewidth}{llRRRRRR}
\toprule
\textbf{Scenario} & \textbf{Protocol} & \rheadcell{\textbf{PDR}\\\textbf{(\%)}} & \rheadcell{\textbf{Mean}\\\textbf{Delay (ms)}} & \rheadcell{\textbf{$p_{95}$}\\\textbf{Delay (ms)}} & \rheadcell{\textbf{Control}\\\textbf{Pkts}} & \rheadcell{\textbf{Control}\\\textbf{MB}} & \rheadcell{\textbf{Recovery}\\\textbf{s (cens./all)}} \\
\midrule
\multirow{7}{*}{\texttt{S6}} & \textbf{INC-SRv6} & \underline{$\mathbf{92.37}$} & $9.60$ & $8.80$ & $116,707$ & $19.51$ & $2.30\text{ s } (0/10)$ \\
 & \textbf{INC-SRv6-uSID} & \underline{$92.31$} & $9.38$ & \underline{$7.28$} & \underline{$115,671$} & \underline{$\mathbf{9.30}$} & \underline{$\mathbf{2.00\text{ s }} (0/10)$} \\
 & AODV-L3 & \underline{$91.25$} & $10.44$ & \underline{$4.22$} & \underline{$177,524$} & $10.31$ & $2.10\text{ s } (0/10)$ \\
 & AntHocNet-L3 & $76.97$ & $30.75$ & $185.38$ & $1,827,881$ & $166.43$ & $2.40\text{ s } (0/10)$ \\
 & OSPF-SRv6 & $56.72$ & $2.50$ & $5.71$ & $39,811$ & $52.75$ & $5.56\text{ s } (1/10)$ \\
 & OLSR-L3 & $29.91$ & $1.20$ & $2.54$ & $33,317$ & $38.57$ & $5.00\text{ s } (1/10)$ \\
 & Q-Routing-L3 & $25.38$ & $2.36$ & $6.05$ & $104,814$ & $6.58$ & $5.78\text{ s } (1/10)$ \\
\midrule
\multirow{7}{*}{\texttt{S7}} & \textbf{INC-SRv6} & \underline{$\mathbf{91.55}$} & $10.24$ & $8.84$ & $121,127$ & $20.27$ & \underline{$\mathbf{2.20\text{ s }} (0/10)$} \\
 & \textbf{INC-SRv6-uSID} & \underline{$91.21$} & $9.90$ & \underline{$7.30$} & \underline{$120,354$} & \underline{$\mathbf{9.69}$} & $2.40\text{ s } (0/10)$ \\
 & AODV-L3 & \underline{$90.25$} & $11.32$ & \underline{$4.35$} & \underline{$179,885$} & $10.47$ & $2.30\text{ s } (0/10)$ \\
 & AntHocNet-L3 & $75.64$ & $32.93$ & $208.07$ & $1,799,634$ & $163.93$ & $3.70\text{ s } (0/10)$ \\
 & OSPF-SRv6 & $55.67$ & $2.46$ & $5.62$ & $39,512$ & $55.00$ & $5.00\text{ s } (1/10)$ \\
 & OLSR-L3 & $30.01$ & $1.20$ & $2.59$ & $33,100$ & $39.92$ & $3.57\text{ s } (3/10)$ \\
 & Q-Routing-L3 & $24.26$ & $2.26$ & $5.87$ & $103,373$ & $6.52$ & $5.75\text{ s } (2/10)$ \\
\midrule
\multirow{7}{*}{\texttt{S10}} & \textbf{INC-SRv6} & \underline{$\mathbf{92.12}$} & $10.57$ & \underline{$8.84$} & \underline{$158,619$} & $26.87$ & \underline{$2.17\text{ s } (0/30)$} \\
 & \textbf{INC-SRv6-uSID} & \underline{$91.78$} & $\mathbf{9.88}$ & \underline{$\mathbf{7.46}$} & \underline{$154,342$} & \underline{$\mathbf{12.51}$} & $2.30\text{ s } (0/30)$ \\
 & AODV-L3 & \underline{$89.12$} & $16.02$ & \underline{$121.26$} & \underline{$247,062$} & \underline{$14.40$} & \underline{$\mathbf{2.07\text{ s }} (0/30)$} \\
 & AntHocNet-L3 & $74.72$ & $40.98$ & $284.32$ & $2,426,164$ & $222.77$ & $2.17\text{ s } (0/30)$ \\
 & OSPF-SRv6 & $39.28$ & $2.19$ & $5.33$ & $43,597$ & $76.43$ & $5.05\text{ s } (11/30)$ \\
 & OLSR-L3 & $20.32$ & $0.98$ & $1.94$ & $37,515$ & $57.63$ & $3.95\text{ s } (9/30)$ \\
 & Q-Routing-L3 & $16.80$ & $2.02$ & $5.73$ & $91,011$ & $5.91$ & $3.95\text{ s } (9/30)$ \\
\bottomrule
\end{tabularx}

\end{table}

Across all six fault-injection scenarios ($\texttt{S5}$--$\texttt{S10}$), \texttt{INC-SRv6} achieves the highest delivery ratio and recovers from $100\%$ of outage events ($0$ censored events across $10$ outage events per scenario in $\texttt{S5}$--$\texttt{S7}$ and $30$ outage events per scenario in $\texttt{S8}$--$\texttt{S10}$):
\begin{itemize}[leftmargin=1.5em]
    \item \textbf{Central Hub Outages ($\texttt{S5}$--$\texttt{S6}$) and Spatial Blackout ($\texttt{S7}$):} When $20$--$40$ central hubs or $40$ contiguous core relays fail simultaneously at $t=15.0$~s, \texttt{INC-SRv6} sustains $92.90\%$ ($\texttt{S5}$), $92.37\%$ ($\texttt{S6}$), and $91.55\%$ ($\texttt{S7}$) PDR, outperforming \texttt{AODV-L3} by $+1.12$ to $+1.54$~pp with $33.1\%$--$36.9\%$ fewer control packets. Because border routers already hold alternative next-hop weights learned from prior suffixes, Stage~4 detours in-flight packets around the void immediately and achieves $2.00$--$2.30$~s cohort recovery.
    \item \textbf{Repeated Relay Churn and High-Speed Stress ($\texttt{S8}$--$\texttt{S10}$):} Under three $4$-s outage waves at $25$~m/s ($\texttt{S8}$--$\texttt{S9}$), \texttt{INC-SRv6} sustains $92.79\%$ and $92.74\%$ PDR with $8.81$~ms $p_{95}$ delay. In the combined stress scenario ($\texttt{S10}$: $35$~m/s flight plus three $40$-relay outage waves), standard \texttt{INC-SRv6} improves PDR from $89.12\%$ (\texttt{AODV-L3}) to $92.12\%$ ($+3.00 \pm 0.90$~pp), reduces mean delay from $16.02$ to $10.57$~ms ($9.88$~ms for \texttt{uSID}), and suppresses $p_{95}$ tail delay by an order of magnitude from $121.26$~ms (\texttt{AODV-L3}) and $284.32$~ms (\texttt{AntHocNet-L3}) to $8.84$~ms ($7.46$~ms for \texttt{uSID}), while recovering across all $30$ outage waves ($2.17$~s mean recovery versus $9$--$11$ censored events for \texttt{OSPF-SRv6}, \texttt{OLSR-L3}, and \texttt{Q-Routing-L3}).
\end{itemize}

\subsection{Control Overhead and 4-Byte Compact \texorpdfstring{\texttt{uSID}}{uSID} Identifier-Width Experiment}
\begin{figure}[H]
    \centering
    \includegraphics[width=0.98\textwidth]{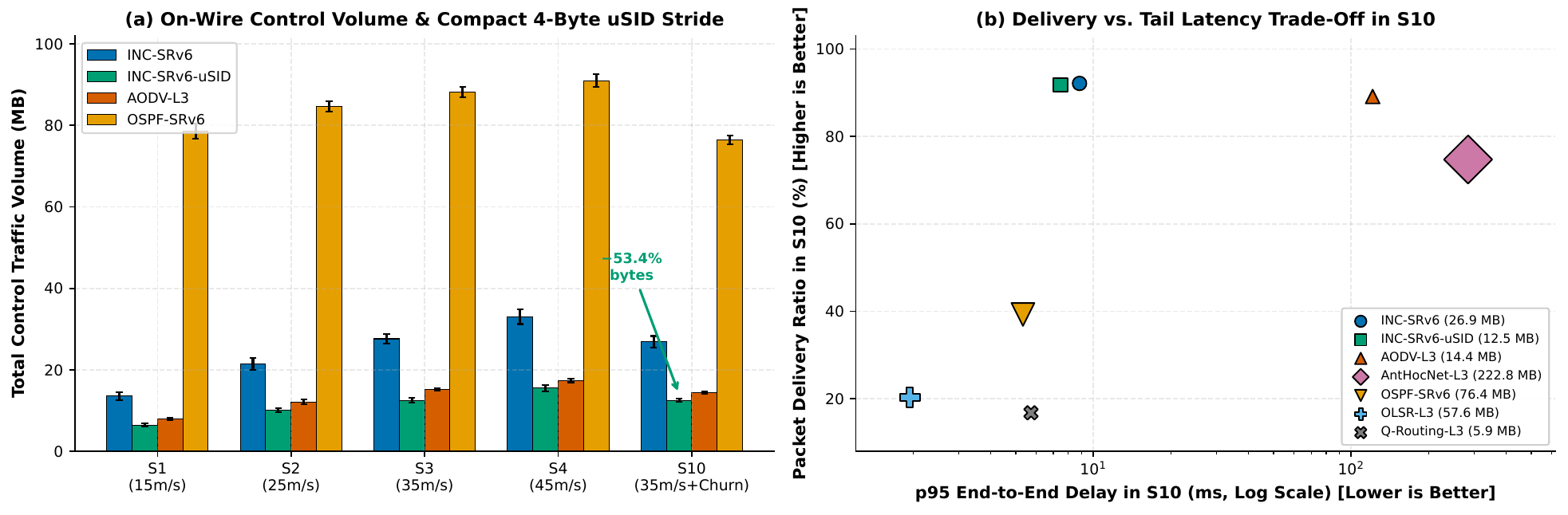}
    \caption{\textbf{Control overhead and 4-byte compact \texttt{uSID} stride model efficiency.} (\textbf{a})~Modeled control byte volume (MB) across $\texttt{S1}$--$\texttt{S4}$ and $\texttt{S10}$, highlighting the $53.4\%$ byte reduction under the $B_{\text{stride}}=4$~B identifier-width model. (\textbf{b})~Pareto trade-off in $\texttt{S10}$ between PDR ($\%$), $p_{95}$ tail delay (ms, log scale), and modeled control byte volume (marker area).}
    \label{fig:overhead}
\end{figure}

Figure~\ref{fig:overhead}a compares the modeled control byte volume across $\texttt{S1}$--$\texttt{S4}$ and $\texttt{S10}$, while Figure~\ref{fig:overhead}b illustrates the three-dimensional Pareto trade-off between delivery ratio, $p_{95}$ tail latency, and control byte volume in $\texttt{S10}$.

In $\texttt{S10}$, 128-bit \texttt{INC-SRv6} sends $35.8\%$ fewer control packets than \texttt{AODV-L3} ($158,619$ versus $247,062$), but incurs higher control byte volume ($26.87$ versus $14.40$~MB) due to $16$-byte SID entries. Adopting the 4-byte compact stride (\texttt{INC-SRv6-uSID}, $B_{\text{stride}}=4$~bytes) reduces control volume in $\texttt{S10}$ by $53.4\%$ to $12.51$~MB ($13.1\%$ below \texttt{AODV-L3}). Across all ten scenarios, \texttt{INC-SRv6-uSID} cuts SRv6 control bytes by $52.1\%$--$54.7\%$ and trims $p_{95}$ delay to $7.28$--$7.59$~ms via shorter serialization times. For a four-entry path (Table~\ref{tab:srh_example}) on a $552$-byte IPv6 \texttt{DATA} packet, SRH encapsulation adds $72$~bytes ($13.0\%$) with 128-bit SIDs and only $24$~bytes ($4.35\%$) with 4-byte \texttt{uSID}, placing \texttt{INC-SRv6-uSID} on the Pareto frontier of high delivery, sub-$10$-ms tail delay, and low overhead (Figure~\ref{fig:overhead}b).

\subsection{Ablation Study of the Four-Stage Workflow}
\label{sec:ablation}
\begin{figure}[H]
    \centering
    \includegraphics[width=0.98\textwidth]{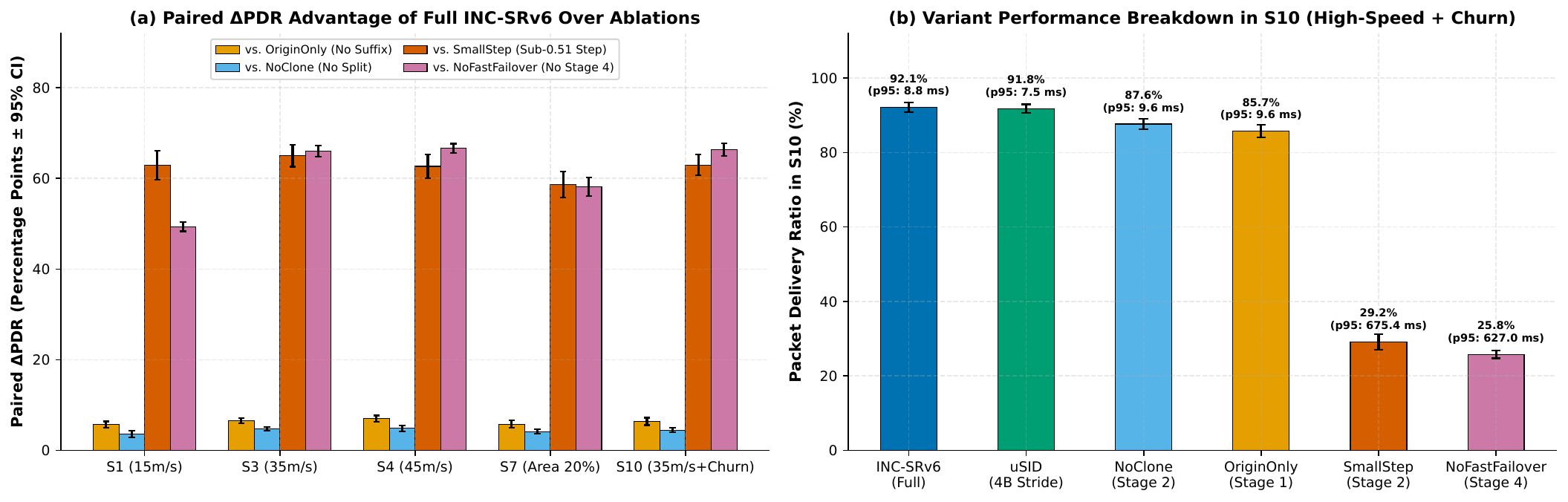}
    \caption{\textbf{Ablation study of the four-stage INC-SRv6 workflow.} (\textbf{a})~Paired $\Delta\text{PDR}$ advantage (percentage points $\pm 95\%$ CI) of full INC-SRv6 over the four mechanism ablations across $\texttt{S1}$, $\texttt{S3}$, $\texttt{S4}$, $\texttt{S7}$, and $\texttt{S10}$. (\textbf{b})~Absolute PDR ($\%$) and $p_{95}$ tail delay (ms) across all six INC-SRv6 variants in $\texttt{S10}$.}
    \label{fig:ablation}
\end{figure}

Figure~\ref{fig:ablation} and Table~\ref{tab:ablation} report paired differences ($\Delta\text{PDR} = \text{PDR}_{\text{Full}} - \text{PDR}_{\text{Ablated}}$ with $95\%$ Student-$t$ CIs and $\Delta\text{Control Bytes}$) between full \texttt{INC-SRv6} and its four mechanism ablations plus the compact-stride \texttt{uSID} variant.

\begin{table}[H]
\caption{Paired ablation study across representative scenarios ($\Delta\text{PDR}$ and $\Delta\text{Control Bytes}$ report paired $\text{INC-SRv6 (Full)} - \text{Comparator}$ differences across 10 paired seeds; negative $\Delta\text{Control Bytes}$ indicates that full INC-SRv6 incurs lower control overhead than the ablated variant; key ablation impacts are underlined).}
\label{tab:ablation}
\centering
\scriptsize
\setlength{\tabcolsep}{3.5pt}
\begin{tabularx}{\linewidth}{@{}lLRRRR@{}}
\toprule
\textbf{Scenario} & \textbf{Ablated Variant} & \rheadcell{\textbf{Variant}\\\textbf{PDR (\%)}} & \rheadcell{\textbf{$\Delta$PDR}\\\textbf{pp $\pm$ 95\% CI}} & \rheadcell{\textbf{Variant}\\\textbf{$p_{95}$ (ms)}} & \rheadcell{\textbf{$\Delta$Control}\\\textbf{Bytes (MB)}} \\
\midrule
\multirow{5}{*}{\texttt{S1}} & \texttt{OriginOnly} & $88.68$ & \underline{$+5.72 \pm 0.67$} & $9.45$ & $-7.80$ \\
 & \texttt{NoClone} & $90.75$ & $+3.65 \pm 0.74$ & $9.15$ & $+1.96$ \\
 & \texttt{SmallStep} & $31.55$ & \underline{$+62.85 \pm 3.18$} & $513.10$ & $-44.56$ \\
 & \texttt{NoFastFailover} & $45.05$ & \underline{$+49.36 \pm 1.03$} & $32.77$ & $+6.90$ \\
 & \texttt{uSID} & $94.70$ & $-0.30 \pm 0.81$ & $7.29$ & $+7.09$ \\
\midrule
\multirow{5}{*}{\texttt{S3}} & \texttt{OriginOnly} & $85.75$ & \underline{$+6.53 \pm 0.60$} & $9.63$ & $-14.64$ \\
 & \texttt{NoClone} & $87.53$ & $+4.75 \pm 0.39$ & $9.58$ & $+4.09$ \\
 & \texttt{SmallStep} & $27.26$ & \underline{$+65.02 \pm 2.44$} & $655.25$ & $-34.19$ \\
 & \texttt{NoFastFailover} & $26.25$ & \underline{$+66.03 \pm 1.20$} & $605.31$ & $+21.42$ \\
 & \texttt{uSID} & $92.33$ & $-0.05 \pm 0.45$ & $7.49$ & $+15.11$ \\
\midrule
\multirow{5}{*}{\texttt{S4}} & \texttt{OriginOnly} & $83.59$ & \underline{$+7.04 \pm 0.67$} & $10.07$ & $-15.78$ \\
 & \texttt{NoClone} & $85.75$ & $+4.88 \pm 0.68$ & $9.77$ & $+4.97$ \\
 & \texttt{SmallStep} & $27.96$ & \underline{$+62.66 \pm 2.58$} & $694.39$ & $-30.02$ \\
 & \texttt{NoFastFailover} & $23.96$ & \underline{$+66.67 \pm 1.00$} & $668.53$ & $+26.94$ \\
 & \texttt{uSID} & $90.39$ & $+0.24 \pm 0.52$ & $7.59$ & $+17.58$ \\
\midrule
\multirow{5}{*}{\texttt{S7}} & \texttt{OriginOnly} & $85.74$ & \underline{$+5.81 \pm 0.84$} & $9.53$ & $-9.95$ \\
 & \texttt{NoClone} & $87.39$ & $+4.16 \pm 0.53$ & $9.39$ & $+3.16$ \\
 & \texttt{SmallStep} & $32.89$ & \underline{$+58.66 \pm 2.83$} & $617.53$ & $-35.19$ \\
 & \texttt{NoFastFailover} & $33.36$ & \underline{$+58.19 \pm 2.08$} & $390.20$ & $+14.10$ \\
 & \texttt{uSID} & $91.21$ & $+0.34 \pm 0.50$ & $7.30$ & $+10.58$ \\
\midrule
\multirow{5}{*}{\texttt{S10}} & \texttt{OriginOnly} & \underline{$85.72$} & \underline{$+6.40 \pm 0.82$} & $9.62$ & \underline{$-14.49$} \\
 & \texttt{NoClone} & $87.64$ & \underline{$+4.48 \pm 0.46$} & $9.64$ & $+3.39$ \\
 & \texttt{SmallStep} & \underline{$29.16$} & \underline{$+62.95 \pm 2.28$} & \underline{$675.39$} & \underline{$-33.63$} \\
 & \texttt{NoFastFailover} & \underline{$25.80$} & \underline{$+66.32 \pm 1.35$} & \underline{$627.01$} & $+20.87$ \\
 & \texttt{uSID} & $91.78$ & $+0.33 \pm 0.30$ & \underline{$7.46$} & \underline{$+14.36$} \\
\bottomrule
\end{tabularx}

\end{table}

The ablation results verify that all four stages are essential and mutually reinforcing:
\begin{enumerate}[leftmargin=1.8em]
    \item \textbf{Stage~1 Full-Suffix Learning (\texttt{OriginOnly}):} Learning only the origin node reduces PDR by $5.10$--$7.04$~pp ($+6.40 \pm 0.82$~pp in $\texttt{S10}$) while \emph{increasing} control bytes by $7.80$--$15.78$~MB (saving $14.49$~MB in $\texttt{S10}$, i.e., $\Delta\text{Control Bytes}=-14.49$~MB), proving that intermediate suffix extraction both pre-populates backup routes and amortizes discovery across flows.
    \item \textbf{Stage~2 Bounded Cloning (\texttt{NoClone}):} Disabling binary splitting ($b_{\max}=0$) reduces PDR by $3.59$--$4.88$~pp ($+4.48 \pm 0.46$~pp in $\texttt{S10}$), as single-branch probes stall in local topological pockets whereas $\le 4$-leaf cloning explores diverse corridors at modest cost ($1.96$--$4.97$~MB).
    \item \textbf{Single-Pass Confidence Calibration (\texttt{SmallStep}):} Using $\alpha(h)=0.35/h < \theta=0.51$ collapses PDR to $27.26\%$--$32.89\%$ ($+62.95 \pm 2.28$~pp gap in $\texttt{S10}$) and inflates $p_{95}$ delay to $513.10$--$694.39$~ms because a single multi-hop \texttt{ACK} ($h \ge 2$, $\alpha(h) \le 0.175$) cannot open the source admission gate from a uniform prior.
    \item \textbf{Stage~4 Fast Failover (\texttt{NoFastFailover}):} Disabling immediate egress-failure recovery drops PDR to $45.05\%$ in $\texttt{S1}$ and $23.96\%$--$25.80\%$ in $\texttt{S4}$/$\texttt{S10}$ ($+66.32 \pm 1.35$~pp in $\texttt{S10}$, $p_{95}$ delay $627.01$~ms), confirming that snapshot-restored local rerouting is the primary driver of high-churn resilience.
\end{enumerate}

\subsection{SRH Processing Pipeline CPU Microbenchmark}
Table~\ref{tab:microbench} reports per-packet CPU execution times on an x86-64 server ($48$ logical CPUs, Python~3.10.13, $7 \times 20,000$ operations via \texttt{time.process\_time\_ns}). Median times are $10.84~\mu\text{s}$ ($92.26$~kpps) for \texttt{ACK} processing (Stages~1 and~3), $33.87~\mu\text{s}$ ($29.52$~kpps) for exploratory forwarding with Stage~2 cloning (Stages~1--3), and $52.24~\mu\text{s}$ ($19.14$~kpps) for link-down failover with snapshot rollback (Stages~1--4), confirming that per-packet SRH processing requires only tens of microseconds even in pure Python.

\begin{table}[H]
\caption{Per-packet CPU execution microbenchmark of the INC-SRv6 processing pipeline on the benchmark host (pure Python~3.10 reference implementation, 7 repeats $\times$ 20,000 packets).}
\label{tab:microbench}
\centering
\footnotesize
\setlength{\tabcolsep}{4.5pt}
\begin{tabularx}{\linewidth}{lLrr}
\toprule
\textbf{Operation} & \textbf{Stages} & \rheadcell{\textbf{Median CPU}\\\textbf{Time (ns/pkt)}} & \rheadcell{\textbf{Pure-Python}\\\textbf{Rate (kpps)}} \\
\midrule
ACK processing & 1 + 3 & $10,839.4$ & $92.26$ \\
Forwarding & 1--3 & $33,870.9$ & $29.52$ \\
Failure branch & 1--4 & \underline{$52,242.3$} & $19.14$ \\
\bottomrule
\end{tabularx}

\end{table}

\section{Discussion and Conclusions}
\label{sec:conclusion}

This paper presented \textbf{INC-SRv6}, an in-network computing mechanism that repurposes the standard IPv6 Segment Routing Header (SRH) and its \texttt{SL} pointer into a distributed route exploration, multi-destination learning, and fast-failover control plane for dynamic UAV networks. By partitioning \texttt{SRH.segment\_list} around the active \texttt{SL} pointer into pre-allocated forward slots and a traversed path suffix, transit routers extract multiple destination/next-hop observations from every arriving packet, gate admission by single-pass confidence ($\theta=0.51$) with bounded binary probing ($\le 4$ leaves), reverse the SRH array at the destination for lookup-free \texttt{ACK} return, and restore pre-transmission SRH snapshots ($\sigma$) for immediate loop-free local repair upon link failure.

Across $1,100$ pure-L3 simulation runs in $200$-node 3D UAV networks, \texttt{INC-SRv6-uSID} matches the delivery of reactive \texttt{AODV-L3} at $15$~m/s cruise speed ($94.70\%$ versus $94.63\%$) while cutting control packets by $41.9\%$ and control bytes by $19.7\%$. Under $35$~m/s high-speed flight with repeated $20\%$ relay churn ($\texttt{S10}$), \texttt{INC-SRv6} improves delivery from $89.12\%$ to $92.12\%$, reduces $p_{95}$ tail delay by an order of magnitude from $121.26$ to $8.84$~ms ($7.46$~ms with \texttt{uSID}), and achieves $100\%$ post-outage cohort recovery, while the 4-byte \texttt{uSID} encoding reduces SRv6 control byte volume by $52.1\%$--$54.7\%$ across all scenarios. In practical wireless deployments, Stage~4 local failover can be gated by a small link-layer retry threshold (e.g., triggering neighbor removal and SRH snapshot rollback after $K$ consecutive short MAC retry failures rather than a single lost frame) to filter transient multipath fading without sacrificing millisecond-scale repair. Future work will integrate cross-layer IEEE~802.11ax/be MAC link-quality hints, implement C-SID wire serialization in eBPF/XDP, and validate INC-SRv6 on a hardware-in-the-loop UAV testbed.

\vspace{6pt}

\authorcontributions{Conceptualization, Z.R.; methodology, Z.R. and Z.L.; software, Z.R.; validation, Z.R. and Z.L.; formal analysis, Z.R.; investigation, Z.R.; data curation, Z.R.; writing---original draft preparation, Z.R.; writing---review and editing, Z.R.; visualization, Z.R. All authors have read and agreed to the published version of the manuscript.}

\funding{This work was supported by the National Key Research and Development Program of China (No.~2024YFE0200302).}

\institutionalreview{Not applicable.}

\informedconsent{Not applicable.}

\dataavailability{All source code for the pure-L3 discrete-event simulator, the INC-SRv6 protocol implementation, the 64 deterministic unit tests, the CPU microbenchmark, and the 1,100-run simulation dataset are openly available at \url{https://github.com/Computing-in-Network/INC-SRv6}.}

\conflictsofinterest{The authors declare no conflicts of interest.}

\begin{adjustwidth}{-\extralength}{0cm}
\reftitle{References}
\externalbibliography{yes}
\bibliography{references}

\begin{thebibliography}{999}

\bibitem[Gupta et~al.(2016)Gupta, Jain, and Vaszkun]{gupta2016survey}
Gupta, L.; Jain, R.; Vaszkun, G.
\newblock {Survey of Important Issues in UAV Communication Networks}.
\newblock {\em IEEE Communications Surveys \& Tutorials} {\bf 2016}, {\em 18},~1123--1152.
\newblock {\url{https://doi.org/10.1109/COMST.2015.2495297}}.

\bibitem[Jiang and Han(2018)]{jiang2018routing}
Jiang, J.; Han, G.
\newblock {Routing Protocols for Unmanned Aerial Vehicles}.
\newblock {\em IEEE Communications Magazine} {\bf 2018}, {\em 56},~58--63.
\newblock {\url{https://doi.org/10.1109/MCOM.2017.1700326}}.

\bibitem[Lakew et~al.(2020)Lakew, Sa'ad, Dao, Na, and Cho]{lakew2020routing}
Lakew, D.S.; Sa'ad, U.; Dao, N.N.; Na, W.; Cho, S.
\newblock {Routing in Flying Ad Hoc Networks: A Comprehensive Survey}.
\newblock {\em IEEE Communications Surveys \& Tutorials} {\bf 2020}, {\em 22},~1071--1120.
\newblock {\url{https://doi.org/10.1109/COMST.2020.2982452}}.

\bibitem[Yao et~al.(2022)Yao, Cheng, Zhang, Zhang, and Zhang]{yao2022uavfleet}
Yao, Z.; Cheng, W.; Zhang, W.; Zhang, T.; Zhang, H.
\newblock {The Rise of UAV Fleet Technologies for Emergency Wireless Communications in Harsh Environments}.
\newblock {\em IEEE Network} {\bf 2022}, {\em 36},~28--37.
\newblock {\url{https://doi.org/10.1109/MNET.001.2100691}}.

\bibitem[Wang et~al.(2024)Wang, Cheng, Zhang, and Liang]{wang2024statqos}
Wang, J.; Cheng, W.; Zhang, W.; Liang, H.
\newblock {Statistical QoS Provisioning Architecture for 6G Satellite-Terrestrial Integrated Networks}.
\newblock {\em Journal of Communications and Information Networks} {\bf 2024}, {\em 9},~34--42.
\newblock {\url{https://doi.org/10.23919/JCIN.2024.10494937}}.

\bibitem[Khan et~al.(2020)Khan, Yau, Noor, and Imran]{oudi2023fanet}
Khan, M.F.; Yau, K.L.A.; Noor, R.M.; Imran, M.A.
\newblock {Routing Schemes in FANETs: A Survey}.
\newblock {\em Sensors} {\bf 2020}, {\em 20},~38.
\newblock {\url{https://doi.org/10.3390/s20010038}}.

\bibitem[Alam and Moh(2022)]{alam2022qlearning}
Alam, M.M.; Moh, S.
\newblock {Survey on Q-Learning-Based Position-Aware Routing Protocols in Flying Ad Hoc Networks}.
\newblock {\em Electronics} {\bf 2022}, {\em 11},~1099.
\newblock {\url{https://doi.org/10.3390/electronics11071099}}.

\bibitem[Zhang et~al.(2023)Zhang, Ma, Zhuang, Xu, Sharma, and Han]{zhang2023qlearninguav}
Zhang, L.; Ma, X.; Zhuang, Z.; Xu, H.; Sharma, V.; Han, Z.
\newblock {Q-Learning Aided Intelligent Routing With Maximum Utility in Cognitive UAV Swarm for Emergency Communications}.
\newblock {\em IEEE Transactions on Vehicular Technology} {\bf 2023}, {\em 72},~3707--3723.
\newblock {\url{https://doi.org/10.1109/TVT.2022.3221538}}.

\bibitem[Filsfils et~al.(2018)Filsfils, Previdi, Ginsberg, Decraene, Litkowski, and Shakir]{rfc8402}
Filsfils, C.; Previdi, S.; Ginsberg, L.; Decraene, B.; Litkowski, S.; Shakir, R.
\newblock {Segment Routing Architecture}.
\newblock RFC 8402, Internet Engineering Task Force (IETF),  2018.
\newblock {\url{https://doi.org/10.17487/RFC8402}}.

\bibitem[Filsfils et~al.(2020)Filsfils, Dukes, Previdi, Leddy, Matsushima, and Voyer]{rfc8754}
Filsfils, C.; Dukes, D.; Previdi, S.; Leddy, J.; Matsushima, S.; Voyer, D.
\newblock {IPv6 Segment Routing Header (SRH)}.
\newblock RFC 8754, Internet Engineering Task Force (IETF),  2020.
\newblock {\url{https://doi.org/10.17487/RFC8754}}.

\bibitem[Filsfils et~al.(2021)Filsfils, Camarillo, Leddy, Voyer, Matsushima, and Li]{rfc8986}
Filsfils, C.; Camarillo, P.; Leddy, J.; Voyer, D.; Matsushima, S.; Li, Z.
\newblock {Segment Routing over IPv6 (SRv6) Network Programming}.
\newblock RFC 8986, Internet Engineering Task Force (IETF),  2021.
\newblock {\url{https://doi.org/10.17487/RFC8986}}.

\bibitem[Ventre et~al.(2021)Ventre, Salsano, Polverini, Cianfrani, Abdelsalam, Filsfils, Camarillo, and Clad]{ventre2020srv6}
Ventre, P.L.; Salsano, S.; Polverini, M.; Cianfrani, A.; Abdelsalam, A.; Filsfils, C.; Camarillo, P.; Clad, F.
\newblock {Segment Routing: A Comprehensive Survey of Research Activities, Standardization Efforts, and Implementation Results}.
\newblock {\em IEEE Communications Surveys \& Tutorials} {\bf 2021}, {\em 23},~182--221.
\newblock {\url{https://doi.org/10.1109/COMST.2020.3036826}}.

\bibitem[Polverini et~al.(2025)Polverini, Cianfrani, Caiazzi, and Scazzariello]{polverini2025srv6detnet}
Polverini, M.; Cianfrani, A.; Caiazzi, T.; Scazzariello, M.
\newblock {SRv6 Meets DetNet: A New Behavior for Low Latency and High Reliability}.
\newblock {\em IEEE Journal on Selected Areas in Communications} {\bf 2025}, {\em 43},~448--458.
\newblock {\url{https://doi.org/10.1109/JSAC.2025.3528806}}.

\bibitem[Secinti et~al.(2018)Secinti, Darian, Canberk, and Chowdhury]{rovira2021sdnuav}
Secinti, G.; Darian, P.B.; Canberk, B.; Chowdhury, K.R.
\newblock {SDNs in the Sky: Robust End-to-End Connectivity for Aerial Vehicular Networks}.
\newblock {\em IEEE Communications Magazine} {\bf 2018}, {\em 56},~16--21.
\newblock {\url{https://doi.org/10.1109/MCOM.2017.1700456}}.

\bibitem[Cianfrani et~al.(2017)Cianfrani, Listanti, and Polverini]{cianfrani2018sdnsrv6}
Cianfrani, A.; Listanti, M.; Polverini, M.
\newblock {Incremental Deployment of Segment Routing into an ISP Network: A Traffic Engineering Perspective}.
\newblock {\em IEEE/ACM Transactions on Networking} {\bf 2017}, {\em 25},~3146--3160.
\newblock {\url{https://doi.org/10.1109/TNET.2017.2731419}}.

\bibitem[Sapio et~al.(2017)Sapio, Abdelaziz, Aldilaijan, Canini, and Kalnis]{sapio2017innetwork}
Sapio, A.; Abdelaziz, I.; Aldilaijan, A.; Canini, M.; Kalnis, P.
\newblock {In-Network Computation is a Dumb Idea Whose Time Has Come}.
\newblock In Proceedings of the 16th ACM Workshop on Hot Topics in Networks (HotNets),  2017, pp. 150--156.
\newblock {\url{https://doi.org/10.1145/3152434.3152461}}.

\bibitem[Ports and Nelson(2019)]{ports2019when}
Ports, D.R.K.; Nelson, J.
\newblock {When Should the Network Be the Computer?}
\newblock In Proceedings of the 17th Workshop on Hot Topics in Operating Systems (HotOS),  2019, pp. 209--215.
\newblock {\url{https://doi.org/10.1145/3317550.3321439}}.

\bibitem[Abdelsalam et~al.(2017)Abdelsalam, Clad, Filsfils, Salsano, Siracusano, and Veltri]{abdelsalam2018srev6}
Abdelsalam, A.; Clad, F.; Filsfils, C.; Salsano, S.; Siracusano, G.; Veltri, L.
\newblock {Implementation of Virtual Network Function Chaining Through Segment Routing in a Linux-Based NFV Infrastructure}.
\newblock In Proceedings of the 2017 IEEE Conference on Network Softwarization (NetSoft),  2017, pp. 1--5.
\newblock {\url{https://doi.org/10.1109/NETSOFT.2017.8004208}}.

\bibitem[Zheng et~al.(2023)Zheng, Ren, Cheng, and Zhang]{zheng2023sfc}
Zheng, S.; Ren, Z.; Cheng, W.; Zhang, H.
\newblock {Performance Analysis for Subgraph Isomorphism Based Embedding Service Function Chains}.
\newblock {\em IEEE Transactions on Network Science and Engineering} {\bf 2023}, {\em 10},~3099--3113.
\newblock {\url{https://doi.org/10.1109/TNSE.2023.3247625}}.

\bibitem[Xhonneux et~al.(2018)Xhonneux, Duchene, and Bonaventure]{lebek2021ebpf}
Xhonneux, M.; Duchene, F.; Bonaventure, O.
\newblock {Leveraging eBPF for Programmable Network Functions with IPv6 Segment Routing}.
\newblock In Proceedings of the 14th International Conference on emerging Networking EXperiments and Technologies (CoNEXT),  2018, pp. 67--72.
\newblock {\url{https://doi.org/10.1145/3281411.3281426}}.

\bibitem[Scarpitta et~al.(2024)Scarpitta, Sidoretti, Mayer, Salsano, Abdelsalam, and Filsfils]{scarpitta2024srv6mon}
Scarpitta, C.; Sidoretti, G.; Mayer, A.; Salsano, S.; Abdelsalam, A.; Filsfils, C.
\newblock {High Performance Delay Monitoring for SRv6-Based SD-WANs}.
\newblock {\em IEEE Transactions on Network and Service Management} {\bf 2024}, {\em 21},~1067--1081.
\newblock {\url{https://doi.org/10.1109/TNSM.2023.3300151}}.

\bibitem[Krishnan(2024)]{rfc9602}
Krishnan, S.
\newblock {Segment Routing over IPv6 (SRv6) Segment Identifiers in the IPv6 Addressing Architecture}.
\newblock RFC 9602, Internet Engineering Task Force (IETF),  2024.
\newblock {\url{https://doi.org/10.17487/RFC9602}}.

\bibitem[Tulumello et~al.(2023)Tulumello, Mayer, Bonola, Lungaroni, Scarpitta, Salsano, Abdelsalam, Camarillo, Dukes, Clad, and Filsfils]{tulumello2023usid}
Tulumello, A.; Mayer, A.; Bonola, M.; Lungaroni, P.; Scarpitta, C.; Salsano, S.; Abdelsalam, A.; Camarillo, P.; Dukes, D.; Clad, F.;  et~al.
\newblock {Micro SIDs: A Solution for Efficient Representation of Segment IDs in SRv6 Networks}.
\newblock {\em IEEE Transactions on Network and Service Management} {\bf 2023}, {\em 20},~774--786.
\newblock {\url{https://doi.org/10.1109/TNSM.2022.3205265}}.

\bibitem[Cheng et~al.(2025)Cheng, Filsfils, Li, Decraene, and Clad]{rfc9800}
Cheng, W.; Filsfils, C.; Li, Z.; Decraene, B.; Clad, F.
\newblock {Compressed SRv6 Segment List Encoding}.
\newblock RFC 9800, Internet Engineering Task Force (IETF),  2025.
\newblock {\url{https://doi.org/10.17487/RFC9800}}.

\bibitem[Atlas and Zinin(2008)]{rfc5286}
Atlas, A.; Zinin, A.
\newblock {Basic Specification for IP Fast Reroute: Loop-Free Alternates}.
\newblock RFC 5286, Internet Engineering Task Force (IETF),  2008.
\newblock {\url{https://doi.org/10.17487/RFC5286}}.

\bibitem[Bashandy et~al.(2025)Bashandy, Litkowski, Filsfils, Francois, Decraene, and Voyer]{rfc9855}
Bashandy, A.; Litkowski, S.; Filsfils, C.; Francois, P.; Decraene, B.; Voyer, D.
\newblock {Topology Independent Fast Reroute Using Segment Routing}.
\newblock RFC 9855, Internet Engineering Task Force (IETF),  2025.
\newblock {\url{https://doi.org/10.17487/RFC9855}}.

\bibitem[Clausen and Jacquet(2003)]{rfc3626}
Clausen, T.; Jacquet, P.
\newblock {Optimized Link State Routing Protocol (OLSR)}.
\newblock RFC 3626, Internet Engineering Task Force (IETF),  2003.
\newblock {\url{https://doi.org/10.17487/RFC3626}}.

\bibitem[Perkins et~al.(2003)Perkins, Belding-Royer, and Das]{rfc3561}
Perkins, C.; Belding-Royer, E.; Das, S.
\newblock {Ad hoc On-Demand Distance Vector (AODV) Routing}.
\newblock RFC 3561, Internet Engineering Task Force (IETF),  2003.
\newblock {\url{https://doi.org/10.17487/RFC3561}}.

\bibitem[Johnson et~al.(2007)Johnson, Hu, and Maltz]{rfc4728}
Johnson, D.; Hu, Y.; Maltz, D.
\newblock {The Dynamic Source Routing Protocol (DSR) for Mobile Ad Hoc Networks for IPv4}.
\newblock RFC 4728, Internet Engineering Task Force (IETF),  2007.
\newblock {\url{https://doi.org/10.17487/RFC4728}}.

\bibitem[Boyan and Littman(1993)]{boyan1993packet}
Boyan, J.A.; Littman, M.L.
\newblock {Packet Routing in Dynamically Changing Networks: A Reinforcement Learning Approach}.
\newblock In Proceedings of the Advances in Neural Information Processing Systems 6 (NIPS),  1993, pp. 671--678.

\bibitem[{Di Caro} et~al.(2005){Di Caro}, Ducatelle, and Gambardella]{dicaro2005anthocnet}
{Di Caro}, G.; Ducatelle, F.; Gambardella, L.M.
\newblock {AntHocNet: An Adaptive Nature-Inspired Algorithm for Routing in Mobile Ad Hoc Networks}.
\newblock {\em European Transactions on Telecommunications} {\bf 2005}, {\em 16},~443--455.
\newblock {\url{https://doi.org/10.1002/ett.1062}}.

\bibitem[da~Costa et~al.(2021)da~Costa, Kunst, and de~Freitas]{da2021qfanet}
da~Costa, L.A.L.F.; Kunst, R.; de~Freitas, E.P.
\newblock {Q-FANET: Improved Q-learning based routing protocol for FANETs}.
\newblock {\em Computer Networks} {\bf 2021}, {\em 198},~108379.
\newblock {\url{https://doi.org/10.1016/j.comnet.2021.108379}}.

\bibitem[Ke et~al.(2024)Ke, Huang, Qiu, Song, Xu, Yin, and Yang]{qiu2024marlfanet}
Ke, Y.; Huang, K.; Qiu, X.; Song, B.; Xu, L.; Yin, J.; Yang, Y.
\newblock {Distributed Routing Optimization Algorithm for FANET Based on Multiagent Reinforcement Learning}.
\newblock {\em IEEE Sensors Journal} {\bf 2024}, {\em 24},~24851--24864.
\newblock {\url{https://doi.org/10.1109/JSEN.2024.3415127}}.

\bibitem[Zhang et~al.(2024)Zhang, Cui, Tso, Li, and Jia]{zhang2024in3}
Zhang, X.; Cui, L.; Tso, F.P.; Li, W.; Jia, W.
\newblock {IN3: A Framework for In-Network Computation of Neural Networks in the Programmable Data Plane}.
\newblock {\em IEEE Communications Magazine} {\bf 2024}, {\em 62},~96--102.
\newblock {\url{https://doi.org/10.1109/MCOM.001.2300587}}.

\bibitem[Kim et~al.(2015)Kim, Sivaraman, Katta, Bas, Dixit, and Wobker]{kim2015int}
Kim, C.; Sivaraman, A.; Katta, N.; Bas, A.; Dixit, A.; Wobker, L.J.
\newblock {In-Band Network Telemetry via Programmable Dataplanes}.
\newblock In Proceedings of the ACM SIGCOMM 2015 Conference (Industrial Demo),  2015, pp. 1--2.

\bibitem[Qian et~al.(2024)Qian, Cui, Tso, Deng, and Jia]{qian2024offsetint}
Qian, M.; Cui, L.; Tso, F.P.; Deng, Y.; Jia, W.
\newblock {OffsetINT: Achieving High Accuracy and Low Bandwidth for In-Band Network Telemetry}.
\newblock {\em IEEE Transactions on Services Computing} {\bf 2024}, {\em 17},~1072--1083.
\newblock {\url{https://doi.org/10.1109/TSC.2023.3323697}}.

\bibitem[Chen et~al.(2024)Chen, Cheng, and Zhang]{chen2024risuav}
Chen, Y.; Cheng, W.; Zhang, W.
\newblock {Reconfigurable Intelligent Surface Equipped UAV in Emergency Wireless Communications: A New Fading-Shadowing Model and Performance Analysis}.
\newblock {\em IEEE Transactions on Communications} {\bf 2024}, {\em 72},~1821--1834.
\newblock {\url{https://doi.org/10.1109/TCOMM.2023.3336223}}.

\bibitem[Coltun et~al.(2008)Coltun, Ferguson, Moy, and Lindem]{rfc5340}
Coltun, R.; Ferguson, D.; Moy, J.; Lindem, A.
\newblock {OSPF for IPv6}.
\newblock RFC 5340, Internet Engineering Task Force (IETF),  2008.
\newblock {\url{https://doi.org/10.17487/RFC5340}}.

\end{thebibliography}
\PublishersNote{}
\end{adjustwidth}

\end{document}